\documentclass[12pt]{iopart}
\usepackage{graphicx}
\usepackage[dvipsnames]{xcolor}
\begin{document}

\title[Turbulent features near the X point..]{Turbulent features near the X point of a DTT-like tokamak plasma}

\author{F. Cianfrani$^1$ \& G. Montani$^{1,2}$}

\address{$^1$ ENEA, Nuclear Department, C. R. Frascati, Via E. Fermi 45, 00044 Frascati (Roma), Italy}
\address{$^2$ Physics Department, ``Sapienza'' University of Rome, P.le Aldo Moro 5, 00185 Roma, Italy}
\eads{\mailto{francesco.cianfrani@enea.it}, \mailto{giovanni.montani@enea.it}}
\vspace{10pt}
\begin{indented}
\item[]July 2024
\end{indented}

\begin{abstract}
The background magnetic geometry at the edge of a tokamak plasma has to be designed in order to mitigate the particle and energy looses essentially due to turbulent transport. The Divertor-Tokamak-Test (DTT) facility under construction at ENEA Frascati will test several magnetic configurations and mitigation strategies, that are usually based on the realization of nontrivial topologies in which one or more X points are present. In order to get a clear understanding of turbulent transport near one of such X points, we perform 3D electro-static fluid simulations of tokamak edge plasma for a DTT-like scenario. We will outline: i) the resulting turbulent spectral features and their dependence on some model parameters (the background pressure gradients and diffusivity) and on the magnetic geometry through a comparative analysis with the results of the companion paper \cite{Cianfrani:24}, ii) the connection between small scale poloidal structures and toroidal asymmetries, iii) the formation of quiescent regions, iv)the crucial role of radial Dirichlet boundary conditions for the excitation of zonal flows that can screen the radial component of the magnetic geometry.    
\end{abstract}

%
%
%
%
%

\section{Introduction}
The geometry of the magnetic field is crucial for energy and particle confinement in a tokamak plasma. The main component is directed toroidally and generated by external (toroidal field) coils, while a secondary poloidal contribution is inductively produced through a discharge in the main plasma column. These two contributions sums up to form an helical magnetic field winding around the main tokamak axis and confining particles in the collisionless limit. However, collisional turbulent transport provides heat and particle looses that must be kept under control to prevent damages of the plasma-facing components (see for instance \cite{Ham:20}). The technical solution is to drive those particles exiting the tokamak core towards some targets capable of sustaining larger heat and particle fluxes than the rest of the chamber walls. This is achieved by realizing nontrivial topological magnetic configurations in the so-called divertor region \cite{Pitcher:97} through some external (poloidal field) coils generating one or more $X$ points where the poloidal magnetic field vanishes.  

The Italian Divertor-Tokamak-Test facility (DTT) \cite{DTT,DTT2} is under construction at ENEA laboratories in Frascati with the aim to test different divertor configurations for ITER and DEMO. The reactor design and the forthcoming experimental campaigns must be supported by an effort to simulate realistic plasma conditions under all the feasible magnetic geometries in order to understand and predict the impact of turbulent transport and the resulting heat and particle fluxes across the targets. The integration of core and scrape-off-layer physics is an hard multi-scale problem requiring extremely costly (gyrokinetic) numerical simulations \cite{Mantica:20,Mishchenko:23}. Instead due to the relatively low temperature and density (the magnetic connection length is much larger than the particle mean free path) the investigation of the edge plasma region alone can be done using Braginskii fluid models \cite{Braginskii:1965} with drift ordering \cite{Simakov:2004} that are the basis of the high-performance turbulence codes for tokamak reactors (as for instance BOUT++\cite{Shanahan:14}, GBS \cite{Halpern:16}, TOKAMAK3X \cite{Tamain:16} and GRILLIX \cite{Stegmeier:18}, see Ref.\cite{Schwander:24} for a recent review). These codes have been validated with experimental data \cite{Oliveira:22,Galassi:22} showing good agreement with the profiles near the midplane and poorer agreement in the divertor region that could be ascribed to the adopted values of dissipative parameters, sheat boundary conditions or to the missing contribution of neutrals, on which the subsequent investigations focused (see for instance Refs.\cite{Giacomin:22,Mancini:24,Quadri:24}). 

In Ref.\cite{Cianfrani:24} we presented a complementary approach to investigate the physics of tokamak edge plasma turbulence: we performed spatially local 3D electro-static simulations for a DTT like scenario varying some model parameters (background pressure gradients and diffusivity) and the magnetic geometry. The resulting spectral features outlined strong dependence on the parameter values, with the limiting case of large (neoclassical) diffusivity and weak background pressure gradients providing a decaying turbulence scenario and the emergence of a sheared radial electric field. In all the examined cases even though axisymmetric $n=0$ modes dominate, 3D transitions from $n\neq 0$ to $n=0$ played a crucial role in establishing the saturated profiles, as anticipated by the pioneering work in Ref.\cite{Biskamp:95}, and the resulting fluxes have the same magnitude as 2D transitions ($n=0\rightarrow n=0$) and the linear contributions (those due to background pressure gradient and the dissipative ones). The inclusion of a poloidal magnetic component breaks the formal isotropy in the poloidal plane of the drift reponse term and the resulting spectra show strong anisotropy with an excess of energy into $m=0$ modes (zonal flow). A similar effect but less intense is also produced by varying the boundary conditions (BCs) along the radial direction from periodic to vanishing Dirichlet ones. Finally, a configuration with a radially sheared poloidal component has been investigated, showing intermediate spectral properties with anisotropic features due to the poloidal drift contribution and average profiles that roughly follow those in SLAB.              

We now extend the analysis in \cite{Cianfrani:24} to a X-point magnetic configuration. The squeeze of the magnetic flux tube near the X-point is expected to break the correlation with the rest of SOL fluctuations, where elongated structures are generated, providing instead ribbon-like structures \cite{Farina:93,Umansky:05} and quiescent regions among them. The presence of such regions in the low field side has been confirmed experimentally \cite{Walkden:17,Walkden:18,Scotti:18,Scotti:20} and derived also in TOKMAK3X \cite{Galassi:17,Galassi:19,Nespoli:19} and GRILLIX \cite{Stegmeier:19} numerical simulations. In \cite{Nem:21} continuous measurements of Langumir probe saturation current from low to high field side SOL at ASDEX upgrade were presented and the spectral analysis revealed the presence of two quiescent regions: one at LFS SOL close to the separatrix and one in the PFR near the separatrix HFS leg. While the LFS quiescent region is characterized by low absolute and relative fluctuations, the PFR one has large relative fluctuations and the physical mechanism behind that is quite obscure.   
		
In this work, we perform the same analysis of Ref.\cite{Cianfrani:24} for the magnetic geometry near the X point and we make a comparative analysis to emphasize the impact of the model parameters and of the magnetic field morphology on the development of toroidaly and poloidally symmetric configurations and on the resulting saturated energy spectra. We will outline: i) the dual role of toroidal transitions, with a direct cascade at long wave-lengths proceeding in connection with an inverse cascade at short wave-lengths; ii) the energy flow due to the drift response that acts as the main source contribution for the electrostatic potential. These two aspects of the flux analysis will be used to explain the observed correlation between short scale poloidal structures and toroidal asymmetries. We will also discuss the different scenarios for periodic and radially vanishing Dirichlet BCs. In the former case, the spectra are quite isotropic and their steepness is strongly dependent on the considered model parameters and different from the results for the magnetic configurations in Ref.\cite{Cianfrani:24}. Moreover, the frequency analysis with spectrograms will reveal the presence of some model dependent quiescent regions below the X point. For neoclassical diffusivity the frequency range is very wide (up to 500KHz) and the quiescent region is in PF region which is both underdense and a local maximum of the electro-static potential. Instead for classical diffusivity quiescence is restricted to two narrow regions to the left of the separatrix legs in correspondence to local fluctuations minima. For vanishing radial Dirichlet BCs, a quite different turbulent scenario will arise with the enhancement of poloidally symmetric $m=0$ modes that provides energy spectra very similar to those obtained in Ref.\cite{Cianfrani:24} for a radially sheared poloidal field, {\it i.e.} ignoring the radial magnetic component of the X point configuration.  

Therefore, some unique features of the physics of turbulence near the X point magnetic geometry are here addressed: 
\begin{itemize}
\item the relevance of the toroidal cascades and of the drift response for establishing nonlinear saturation, 
\item the dependence of the spectral features on the model parameters, that could be relevant in connection with turbulence measurements to test some of the assumptions in the modelization of a realistic tokamak edge plasma, 
\item the formation of quiescent regions due to a local reduction of fluctuations or to the background field dynamics,   
\item the development of zonal flows in conjunction with the mathematical assumption of radial vanishing Dirichlet BCs and their role in hiding the radial magnetic field component effectively spoiling the full X-point morphology.  
\end{itemize}

The organization of the work is as follows. In section \ref{sec2} the model equations and the numerical scheme are presented, outlining the physical assumptions, the morphology of the magnetic geometry and the adopted technical tools. The results of the simulations are shown in section \ref{sec3}, with the degree of toroidal and poloidal symmetry, the energy spectra and the analysis of fluxes discussed  for periodic and vanishing Dirichlet BCs for the radial coordinate in subsections \ref{sec3.1} and \ref{sec3.2}, respectively. Brief concluding remarks follow in section \ref{sec4}.      

\section{Basic equations and methodology}\label{sec2}
We consider the fluid description of edge tokamak plasma in the low frequency limit ($\omega<<\Omega_i$, $\Omega_i$ being ion Larmor frequency) with drift-ordering approximation assuming equal and constant electron and ion temperatures $T_e=T_i=T$. Neglecting parallel velocity and magnetic fluctuations (electro-static limit), drift turbulence can be described in terms of two variables, the electrostatic potential $\phi$ and the pressure $p=p_0+\tilde{p}$, which we split into a background component $p_0$ accounting for the background triggering for linear instability and the fluctuation $\tilde{p}$. The model is derived from the conditions of charge neutrality and energy balance \cite{Beyer:98,Montani:23} neglecting magnetic drifts and the compressibility of ion diamagnetic velocity, resulting in the following system of equations 
\begin{equation}
\left\{\begin{array}{c} 
\partial_t \Delta_\bot \phi + \{\phi,\Delta_\bot \phi\} = \displaystyle\frac{BF}{\eta}\,\nabla_\parallel^2 (\tilde{p}-\phi)  + \mu \,\Delta_\bot^2\phi\hspace{.5cm}\\
\partial_t \tilde{p} + \{\phi,\tilde{p}\} = -\hat{k}\,\delta_{y} \phi + \displaystyle\frac{DF}{\eta}\,\nabla_\parallel^2 (\tilde{p}-\phi) + \chi_\bot\, \Delta_\bot\tilde{p}  
\end{array}
\right.\label{ES}
\end{equation}
The two equations above are for vorticity $\Delta_\bot \phi$ and perturbed pressure $\tilde{p}$. They have a similar structure with the nonlinear contributions due to the $\mathrm{E \times B}$ advection expressed in terms of some Poisson brackets (see below) at the left-hand side, while the right-hand side contains the drift response and the dissipative contributions, written in terms of the perpendicular Laplacian $\Delta_\bot$ and the squared parallel gradients $\nabla_\parallel$ with respect to the magnetic field. For a generic magnetic configuration, these operators can be written in terms of the magnetic unit vector $\hat{b}=\vec{B}/|\vec{B}|$ as 
\begin{equation}
\nabla_\parallel=\hat{b}\cdot\vec{\partial}\qquad \Delta_\bot=\Delta-(\nabla_\parallel)^2\,.
\end{equation}
The only formal difference between the two equations in (\ref{ES}) is due to the first term on the right-hand side of the second one: it is due to the Poisson brackets $\{\phi,p_0\}$ and its coefficient $\hat{k}$ is proportional to the inverse background pressure gradient length $1/\ell_0=\partial_xp_0/p_0$. A reduced model with one single equation has been considered in Refs.\cite{Montani:23,Carlevaro:23} (see also \cite{Montani:22}) neglecting this contribution and identifying the dissipative parameters. Instead we will discuss in this work the relevance of $\ell_0$ for the resulting drift turbulent scenario close to the X point.

The Poisson brackets are given in terms of some generalized derivative operators $\delta_x$ and $\delta_y$ living in the perpendicular plane to the magnetic field
\begin{equation}
\{f,g\}=\delta_{x} f\,\delta_{y} g - \delta_{x} g\,\delta_{y} f  
\end{equation}
where $\delta_{x}$ and $\delta_{y}$ read 
\begin{eqnarray}
\delta_{x}=\frac{1}{\sqrt{b_y^2+b_z^2}}\left[(1-b_x^2)\partial_x-b_x(b_y\partial_y+b_z\partial_z\right] \\
\delta_{y}=\frac{1}{\sqrt{b_y^2+b_z^2}}\left(b_z\partial_y - b_y\partial_z\right)
\end{eqnarray}
 
\begin{figure}[h]%
\centering
\includegraphics[width=.7\textwidth]{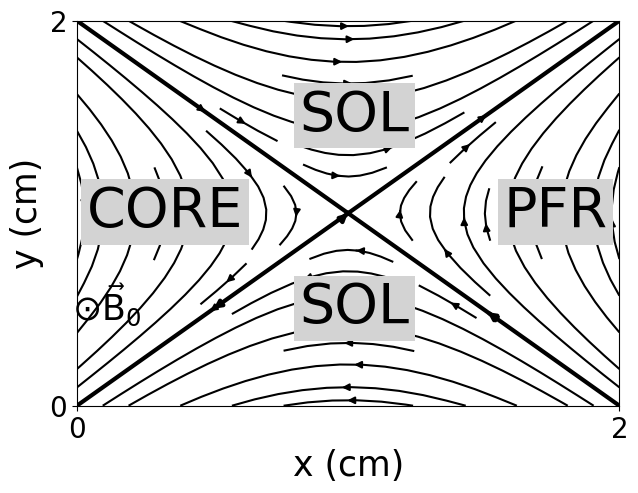}
\caption{The poloidal magnetic field lines of the considered X point configuration. The SOL, PF and core regions are indicated within the simulation domain. The (forward) direction of the toroidal magnetic field $\vec{B}_0$ is also shown and it is aligned with the induced plasma current.}
\label{fig_mgeo}
\end{figure}

In what follows, we take some coordinates $x$ and $y$ in the poloidal plane that are close to the X point (they extend over a 2cm$\times$2cm square centered at the X point position) and correspond to radial and poloidal directions, respectively, while the third coordinate $z$ covers the whole toroidal axis. We neglect toroidicity, so the resulting system of coordinates is effectively a Cartesian one and all the effects we will discuss are entirely due to the X point magnetic geometry, which we choose as follows 
\begin{equation}
B_x/B_0=\alpha\,y\qquad B_y/B_0=\alpha\,x\qquad B_z=B_0\,,\label{mag_field}
\end{equation}
with $\alpha=1.5\times 10^{-3}$ and the field lines in the poloidal plane are shown in Fig.\ref{fig_mgeo}.

Viscosity $\mu$, resistivity $\eta$ and the additional parameters $B$, $D$ and $F$ are fixed according with Braginskii values and the chosen normalization. In particular, poloidal coordinates $x$ and $y$ and the toroidal one $z$ are rescaled so that the corresponding dimensionless values range in $[-5,5]$, while electro-static potential and pressure are normalized with $e/K_BT$, $K_B$ being the Botzmann constant, and $n_0K_BT$, respectively, for a typical DTT-like scenario \cite{DTT} with
\begin{equation}
n_0=5\cdot 10^{13}cm^{-3}\quad T= 100eV,\quad B_0=3T\quad R=2.14m\quad a=68cm.
\end{equation}

The perpendicular diffusivity $\chi_\bot$ and the background pressure gradient length $\ell_0$ are the two parameters that are varied to test the sensitivity of the model. In particular, the classical Braginskii value $\chi_{\mathrm{class}}=110 cm^2/s$ and a value ten times larger $\chi_{\mathrm{neocl}}=10\chi_{\mathrm{class}}$, that is representative of the neoclassical value, are considered for $\chi_\bot$, while two cases are discussed for $\ell_0$: strong $\ell_0=2cm$ and weak $\ell_0=10cm$ gradients.  

The numerical integration scheme is discussed with details in Ref.\cite{Cianfrani:24}. We adopt the explicit 4$^\mathrm{th}$ order Runge-Kutta method in Fourier space. At each step, vorticity and perturbed pressure are advanced, while the electro-static potential is calculated through the numerically pre-comupted inverse Laplacian operator. Extensive use is made of Fast Fourier Transforms (FFTs) to implement the pseudospectral method for all those (nonlinear) terms involving the product of two or three fields, such as $\mathrm{E\times B}$ advection and drift response contributions. Half of the modes are zero-padding to avoid aliasing error and we use  
 $\mathrm{56\times 56\times 24}$ modes covering physical length scales down to 3 times the Larmor radius (for $l,m=\pm 14$, $l$ and $m$ being radial and poloidal mode numbers). The simulations are run in python and the basic adopted packages are \emph{scipy} and \emph{numpy}. 

In what follows, we implement the model equations (\ref{ES}) in two ways: 
\begin{itemize}
\item the standard formulation in FS resulting in fully periodic boundary conditions along all the coordinates,
\item a nontrivial formulation for $x$ simulation domain with sine transform and a background configuration that is symmetric under the reflection over the plane passing through the midpoint, so obtaining skew-symmetric fields under the same reflection and realizing vanishing Dirichlet boundary conditions on half of the domain (see Ref.\cite{Cianfrani:24} for a more detailed explanation).   
\end{itemize}

\begin{figure}[h]%
\centering
\includegraphics[width=\textwidth]{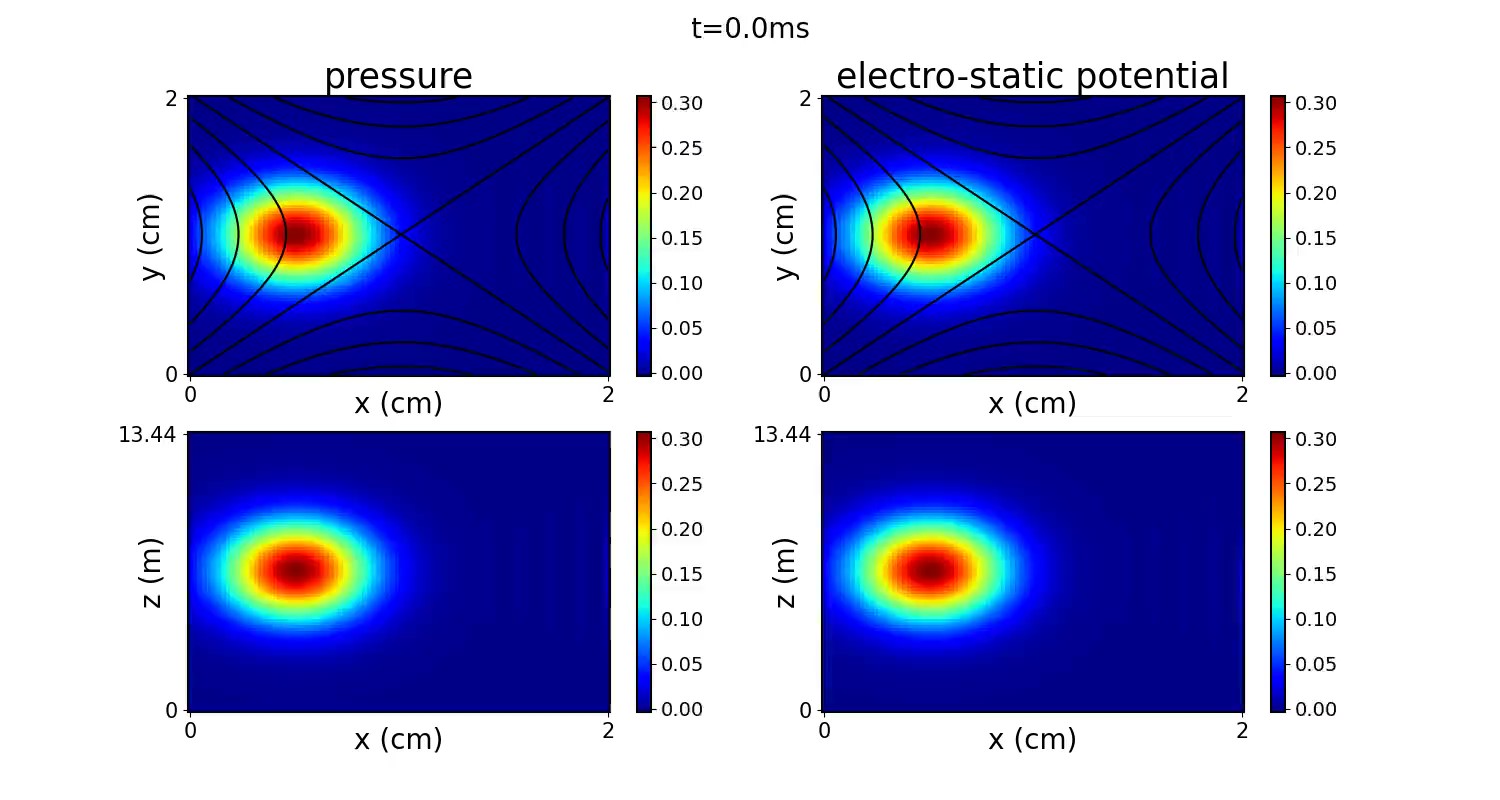}
\caption{The initial condition of the simulations are the same 3D Gaussian function for normalized $\tilde{p}$ (left) and $\phi$ (right). The top figures are for the poloidal sections, $(x,y)$ plane, and the bottom figures are for the radial-toroidal $(x,z)$ plane.}
\label{fig_init}
\end{figure}

The initial conditions are shown in Fig.\ref{fig_init}: they are 3D Gaussian wave packets centered inside the simulation domain with $\phi=\tilde{p}$ minimizing the initial drift response. The height and width are chosen such that turbulence is self-sustained \cite{Scott:90} and no linear phase occurs.

\section{Results}\label{sec3}

The simulation results are presented in this section for the four considered cases (a)$\chi_\bot=\chi_{neocl}$ and $\ell_0=2cm$ in red, b)$\chi_\bot=\chi_{class}$ and $\ell_0=2cm$ in green, c)$\chi_\bot=\chi_{neocl}$ and $\ell_0=10cm$ in blue, d)$\chi_\bot=\chi_{class}$ and $\ell_0=10cm$ in yellow) with fully periodic and vanishing radial Dirichlet boundary conditions. It is performed a comparative analysis of the energy spectra defined as  
\begin{equation}
E_\phi(\vec{k})=\frac{1}{2}\,|\nabla_\bot\phi|^2(\vec{k})\qquad E_{\tilde{p}}(\vec{k})=\frac{1}{2}|\tilde{p}(\vec{k})|^2
\end{equation}
 and of the corresponding fluxes, {\it i.e.}
\begin{equation}
\left\{\begin{array}{c} 
\partial_t E_\phi = T_\phi + \Gamma_{\mathrm{drift},\phi}  - \Gamma_{\mathrm{visc}}\hspace{.5cm}\\
\partial_t E_{\tilde{p}} = T_{\tilde{p}} + \Gamma_{\hat{k}} + \Gamma_{\mathrm{drift},\tilde{p}} - \Gamma_{\mathrm{diff}}   
\end{array}
\right.\label{EES}
\end{equation}
whose definition is given in Ref.\cite{Cianfrani:24}.

\subsection{Fully periodic boundary conditions}\label{sec3.1}

The multi-panel figure Tab.\ref{frames_X} shows the profiles of $\tilde{p}$ and $\phi$ at a given time $t=0.335\mathrm{ms}$ (in normalized units $\tau=\Omega_i t= 50000$ with the simulation time step $d\tau=0.1$). The two upper panels describe the cases with strong background pressure gradient ($\ell_0=2\mathrm{cm}$) and the lower ones are for weak gradients ($\ell_0=10\mathrm{cm}$) with neoclassical/classical diffusivity coefficients $\chi_\bot$ to the left/right. In each panel the figures above correspond to the poloidal plane $(x,y)$ (where magnetic field lines are also plotted) and the figures below are for the radial-toroidal plane $(x,z)$.

\begin{table}
\begin{tabular}{|c|c|}

      \hline \\
            \includegraphics[trim={3.5cm 1.5cm 3.5cm 1.7cm},clip,width=80mm]{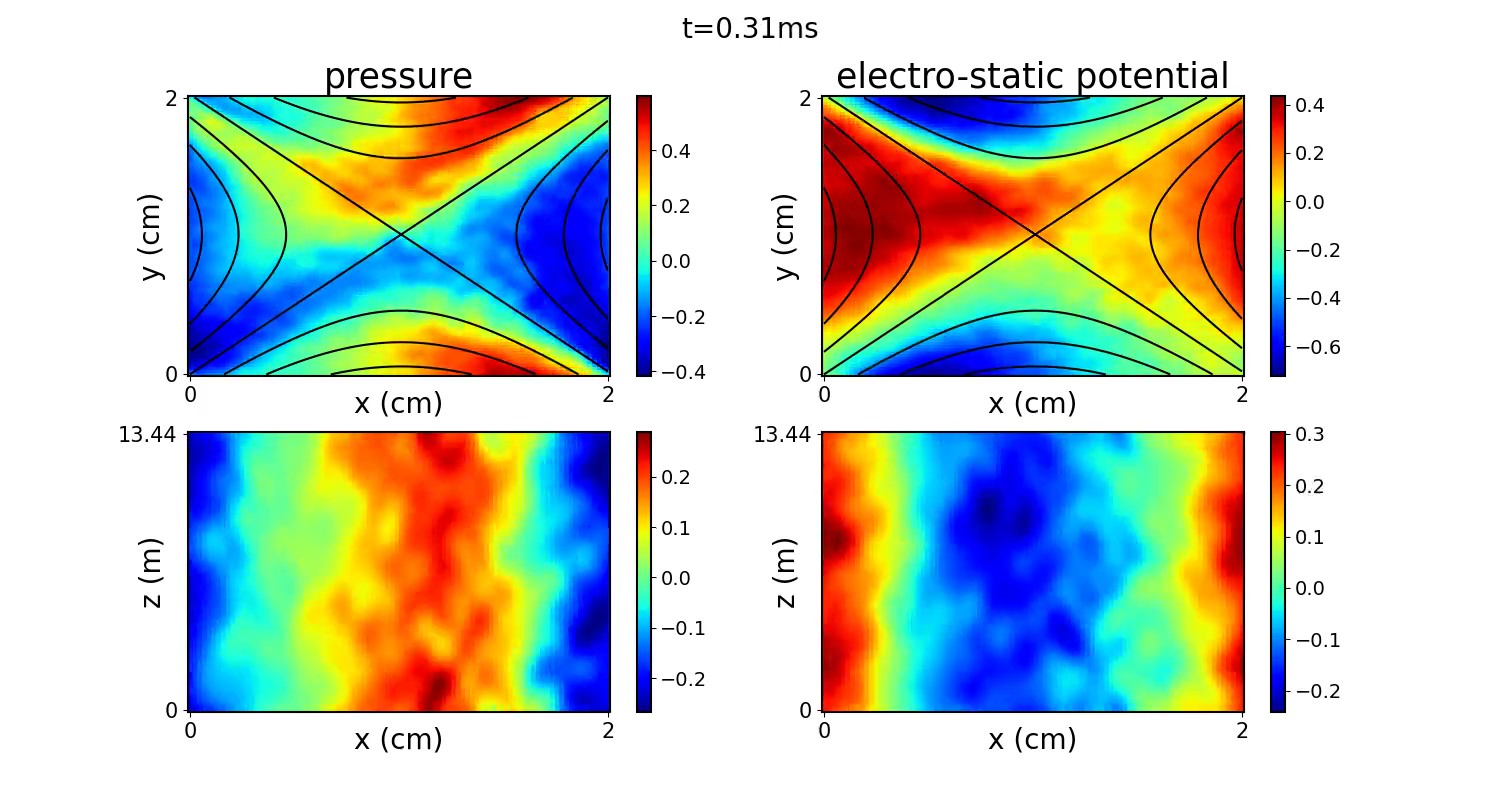} & \includegraphics[trim={3.5cm 1.5cm 3.5cm 1.7cm},clip,width=80mm]{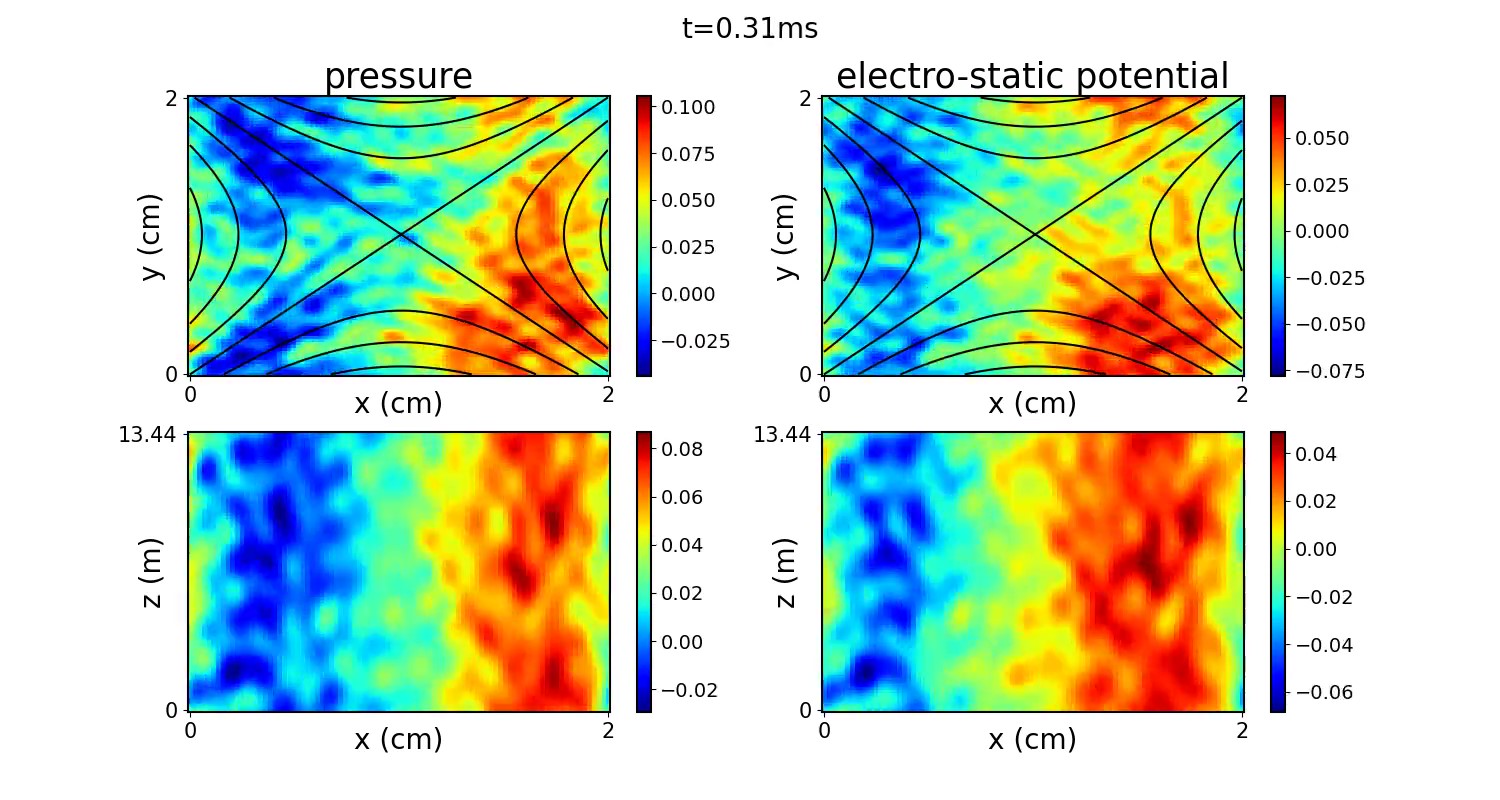} \\
			{\color{red}\textbullet} a) $\chi_{\mathrm{neocl}}$ $\ell_0$=2cm  & {\color{ForestGreen}\textbullet} b) $\chi_{\mathrm{class}}$ $\ell_0$=2cm \\
			\hline \\
      \includegraphics[trim={3.5cm 1.5cm 3.5cm 1.7cm},clip,width=80mm]{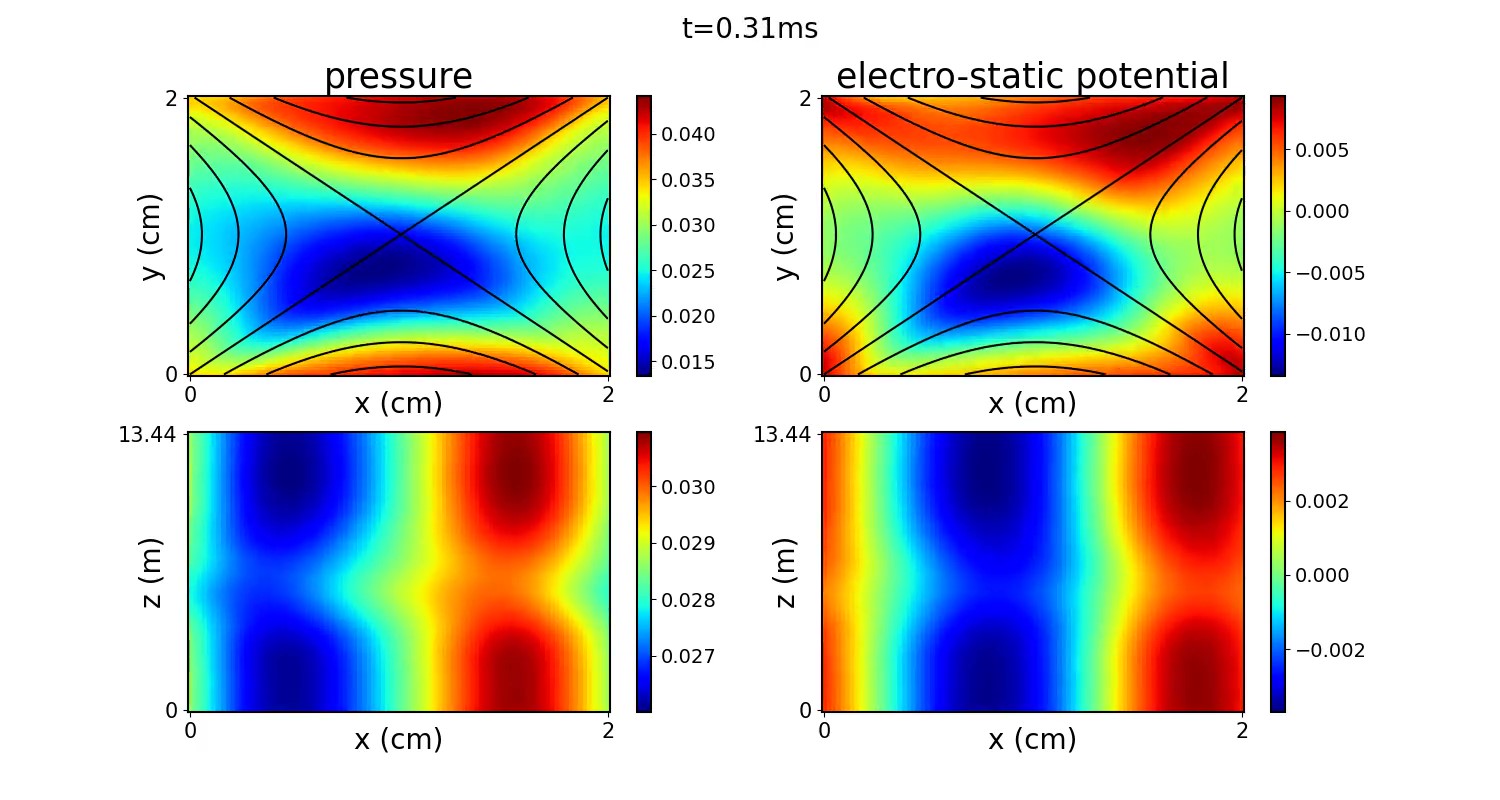} & \includegraphics[trim={3.5cm 1.5cm 3.5cm 1.7cm},clip,width=80mm]{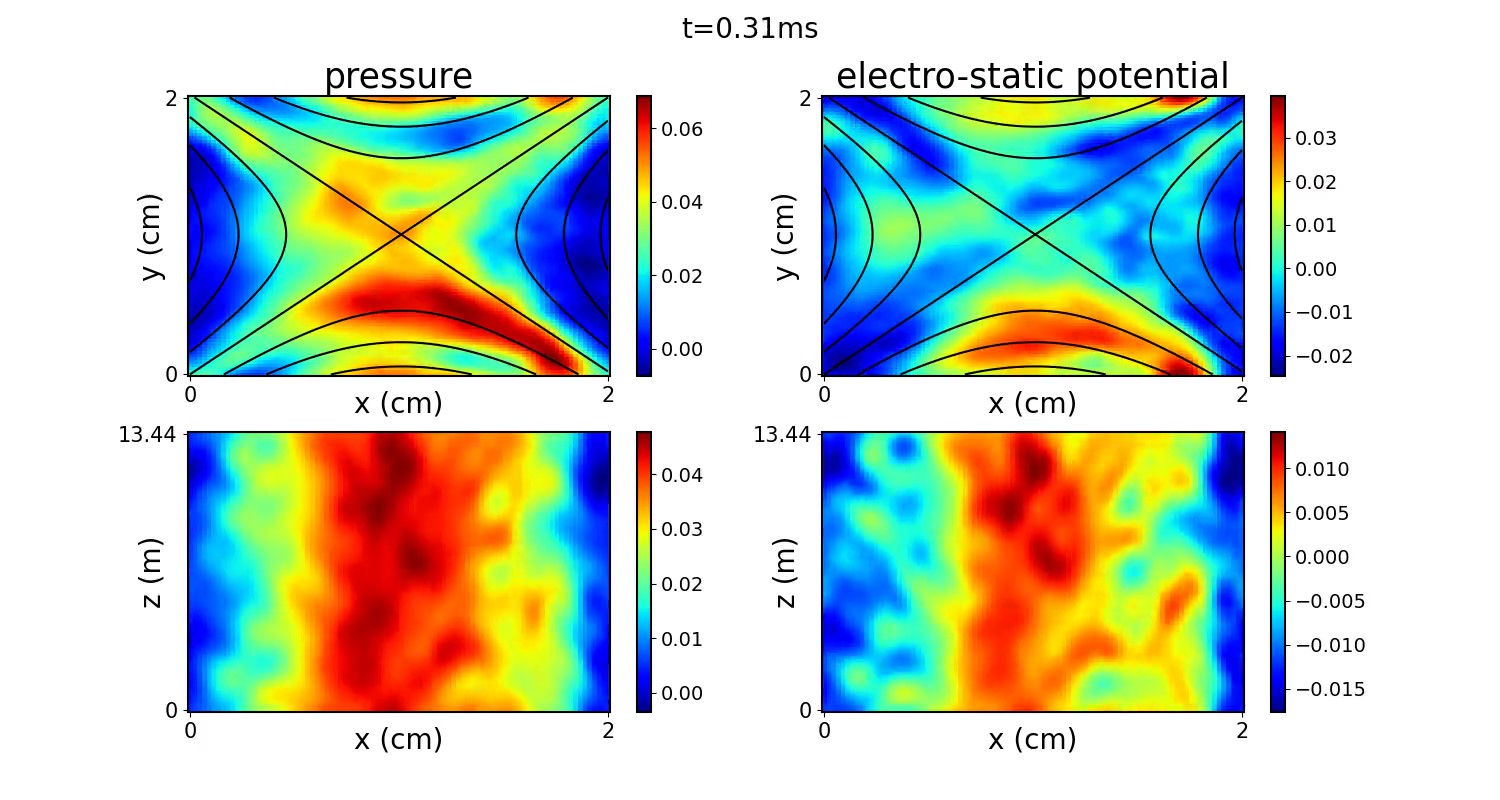} \\
			{\color{blue}\textbullet} c) $\chi_{\mathrm{neocl}}$ $\ell_0$=10cm & {\color{Goldenrod}\textbullet} d) $\chi_{\mathrm{class}}$ $\ell_0$=10cm \\			
			\hline 
\end{tabular}
\caption{The multi-panel figure showing the fields profile at fixed time $t=0.31ms$ for the simulations with fully periodic BCs. Each panel is for one choice of the model parameters: a) $\chi_\bot=\chi_{\mathrm{neocl}}$ and $\ell_0$=2cm (top left), b) $\chi_\bot=\chi_{\mathrm{class}}$ and $\ell_0$=2cm  (top right), c) $\chi_\bot=\chi_{\mathrm{neocl}}$ and $\ell_0$=10cm (bottom left), d) $\chi_\bot=\chi_{\mathrm{class}}$ and $\ell_0$=10cm  (bottom right). In each panel the poloidal (top) and the radial-toroidal (bottom) sections of $\tilde{p}$ (left) and $\phi$ (right) are plotted with the corresponding colorbars next to each plot.}
\label{frames_X}
\end{table}

It is worth noting that by decreasing diffusivity (left to right) and/or increasing $\ell_0$ (top to bottom) one sees a lower degree of toroidal symmetry and smaller vortex structures in the poloidal plane, in qualitative agreement with the results of Ref.\cite{Cianfrani:24} obtained with different magnetic configurations (SLAB, constant and radially sheared poloidal magnetic field).  

\begin{figure}[h]
\centering
\includegraphics[width=\textwidth]{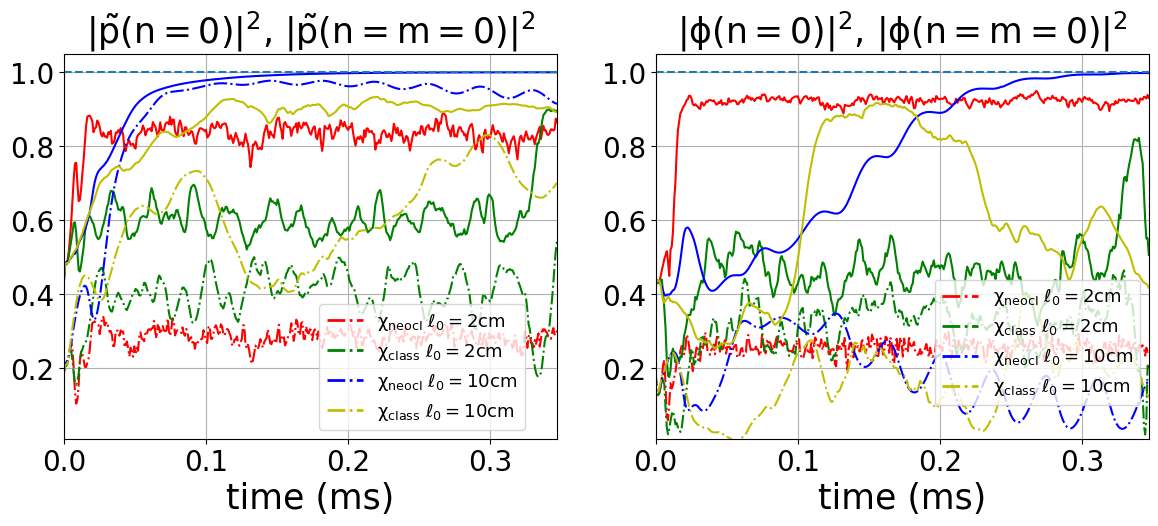}
\caption{The ratios of $n=0$ (solid lines) and of $n=m=0$ (dash-dotted lines) mode amplitudes over the total amplitude for $\tilde{p}$ (left) and $\phi$ (right) vs time (in ms) for the four considered cases with fully periodic BC.}
\label{modes_Xzf}
\end{figure}

The fractions of $n=0$ (solid lines) and $n=m=0$ (dash-dotted lines) modes vs time are plotted in figure \ref{modes_Xzf}. Saturated values for $\tilde{p}$ (left) solid lines are reached quite fast and range from practically $1$ (full toroidal symmetry) for the case (blue) with $\chi_\bot=\chi_{neocl}$ and $\ell_0=10cm$ to $\sim 60\%$ for $\chi_\bot=\chi_{class}$ and $\ell_0=2cm$ (green). The same curves for $\phi$ show a more complex behavior for weak gradients with longer time to saturation (blue) and a sort of fast decay (yellow) that could be part of an oscillation. In all cases, the degree of toroidal symmetry stays quite high, with more that 50\% of modes being axisymmetric.

\begin{figure}[h]
\centering
\includegraphics[width=\textwidth]{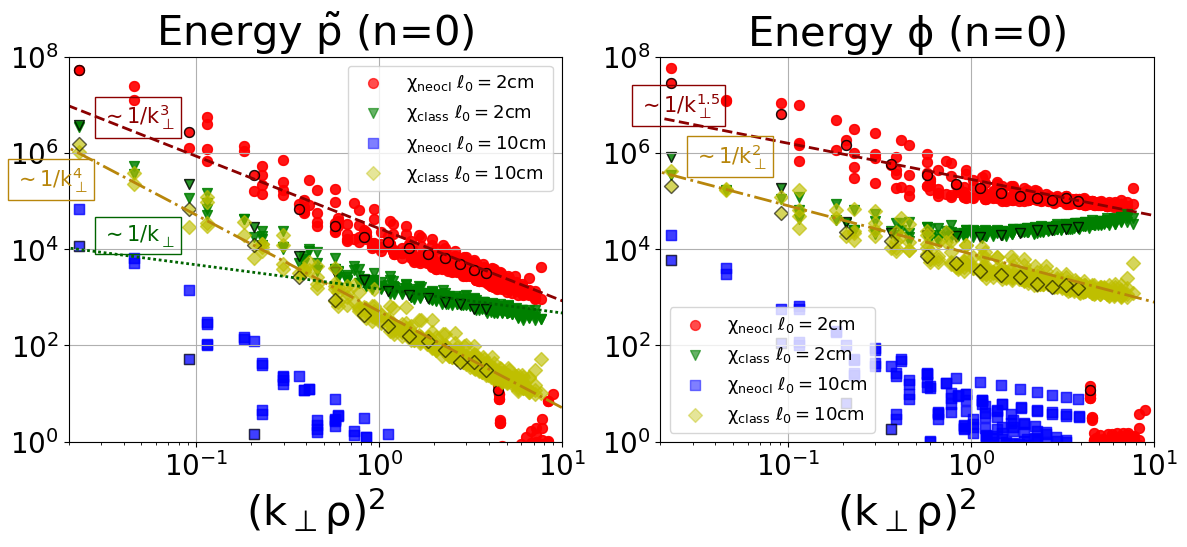}
\caption{The scatter plots of normalized axisymmetric spectral energies $E_{\tilde{p}}$ (left) and $E_\phi$ (right) vs $(k_\bot\rho)^2$ for the four considered cases with fully periodic BCs. They are computed by averaging over a time window of $\sim 0.26ms$. Some curves are also plotted that approximate quite well some of the spectra profiles and which behave as: $1/k_\bot^3$ (dashed dark red line), $1/k_\bot^4$ (dash-dotted dark yellow line) and $1/k_\bot$ (dotted dark green line) for $E_{\tilde{p}}$ (left) and $1/k_\bot^{1.5}$ (dashed dark red line) and $1/k_\bot^2$ (dash-dotted dark yellow line) for $E_\phi$ (right).}
\label{en_spectra_X}
\end{figure}

The dash-dotted lines reach lower saturated values signaling that zonal flows are present but they are not leading the dynamics. This is confirmed by the scatter plot of $n=0$ energy spectra in Fig.\ref{en_spectra_X} in which $m=0$ modes are outlined with black edges and they do not reach highest values than the other points with the same $k_\bot$. For a comparison one can look at the same plots for constant and radially sheared poloidal magnetic fields in Ref.\cite{Cianfrani:24} (see Figs. 9 and 20), where $m=0$ modes stand out from other spectral points. The reason for that was found in the drift term breaking the formal symmetry between the coordinates in the poloidal plane, resulting in an effective energy pumping to $m=0$ modes. On the contrary, the drift response due to the X point configuration does not break the formal symmetry between $x$ and $y$ and it does not produce any enhancement of $m=0$ modes.   

Although the fluctuations at fixed $k_\bot$ are not completely negligible, one can approximate the spectra as being isotropic and use the reference curves in figure. This way, ones sees how the profiles become steeper by increasing diffusivity (from green to red) and/or decreasing background pressure gradients (from green to yellow), while in the blue case they are not even stationary and become more and more stiff with a decaying turbulence scenario. Despite these qualitative similarities with the profiles in Ref.\cite{Cianfrani:24}, in particular with those in SLAB (see Fig.4), the logarithmic spectral steepness parameters are very different: while $\tilde{p}$ spectra go like $1/k_\bot^4$ and $1/k_\bot^2$ in the two SLAB cases with $\ell_0=2cm$ ($\chi_\bot=\chi_{neocl}$ (red) and $\chi_\bot=\chi_{class}$ (green)) and $1/k_\bot^3$ for $\chi_\bot=\chi_{class}$ and $\ell_0=10cm$ (yellow), the same curves from Fig.\ref{en_spectra_X} are $1/k_\bot^3$ (red), $1/k_\bot$ (green) and $1/k_\bot^4$ (yellow), respectively. Similarly for $\phi$ the spectra in SLAB are $\sim 1/k_\bot^3$ (red), $\sim 1/k_\bot$ (green) and $\sim 1/k_\bot^2$ (yellow) and the same here go as $1/k_\bot^{1.5}$ (red), $const.$ (green) and $1/k_\bot^2$ (yellow). We see how just $\phi$ spectral profiles for case d) (yellow) is on average the same as in SLAB, while the X point magnetic geometry generically provides more gentle profiles.

This findings outline how the local field spectral profiles are strongly correlated with both the model parameters and the morphology of the X point magnetic geometry.

In order to distinguish toroidal from poloidal cascades the nonlinear transfer functions $T_{\tilde{p}}$ and $T_\phi$ in Eq.(\ref{EES}) can be split into toroidal transitions from $n \neq 0$ to $n=0$ and 2D ones within $n=0$ modes:
\begin{equation}
T_{\mathrm{..}} = T^{(n\neq0\rightarrow n=0)}_{\mathrm{..}} +  T^{(2D)}_{\mathrm{..}}\,. 
\end{equation} 

\begin{figure}[h]
\centering
\includegraphics[width=\textwidth]{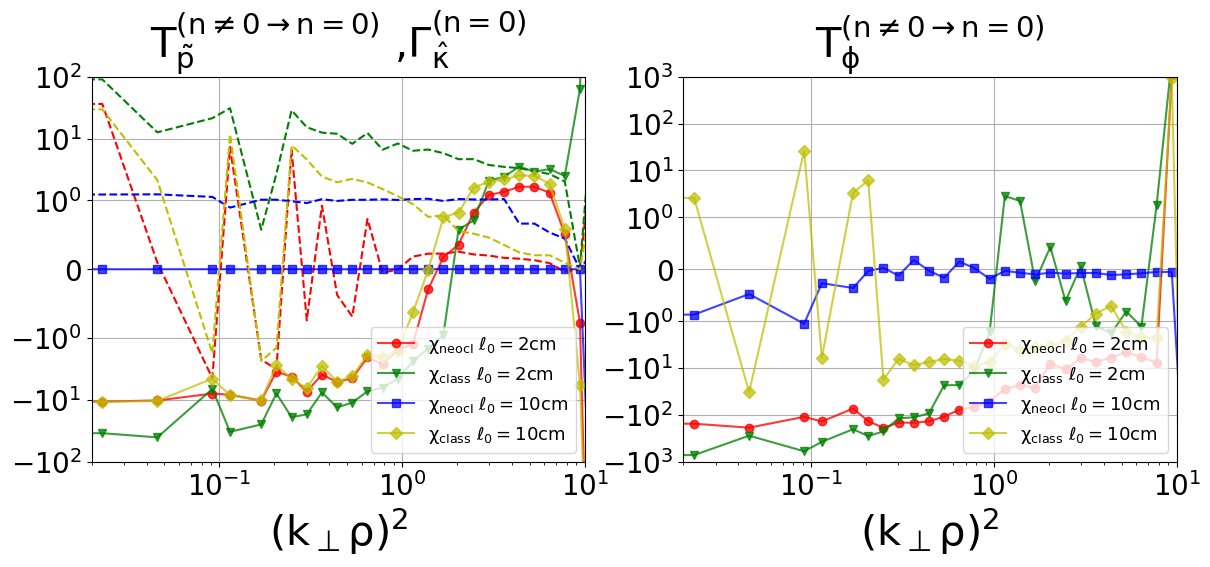}
\caption{The nonlinear transfer functions of toroidal transitions $n\neq0\rightarrow n=0$ (solid lines) for $\tilde{p}$ (left) and $\phi$ (right) vs $(k_\bot\rho)^2$ and the flux due to background pressure gradient (dashed lines, left) for the four considered cases with fully periodic BCs. Each plotted value is normalized with the corresponding dissipative flux, {\it i.e.} the diffusivity and the viscous contributions for $\tilde{p}$ and $\phi$, respectively. The chosen values of $(k_\bot\rho)^2$ are constructed from the computational Cartesian grid in FS by taking the five smallest values of $k_\bot$ and by logarithmically sampling the relic part.}
\label{Tpphi_X}
\end{figure}

$T^{(n\neq0\rightarrow n=0)}_{\tilde{p}}$ (left) and $T^{(n\neq0\rightarrow n=0)}_{\phi}$ (right) are shown in Fig.\ref{Tpphi_X} by normalizing with the corresponding dissipative terms, {\it i.e.} the diffusive $\Gamma_{\mathrm{diff}}$ and viscous $\Gamma_{\mathrm{visc}}$ fluxes, respectively, and using a logarithmic sampling for $(k_\bot\rho)^2$. 

\begin{figure}[h]
\centering
\includegraphics[width=\textwidth]{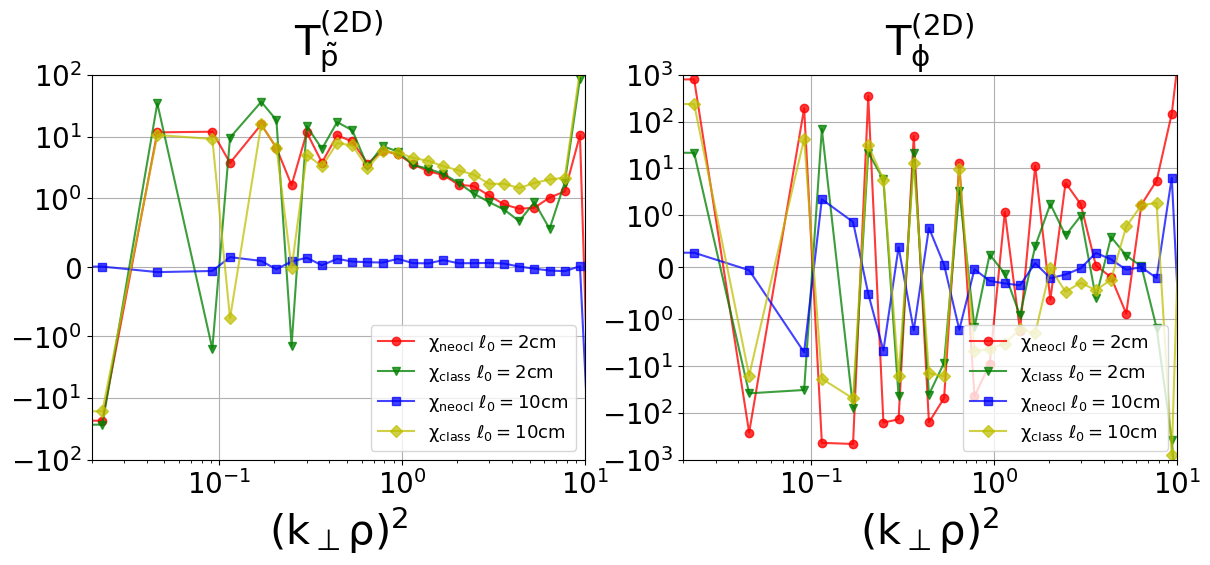}
\caption{The nonlinear transfer functions of 2D transitions $n=0\rightarrow n=0$ (solid lines) for $\tilde{p}$ (left) and $\phi$ (right) vs $(k_\bot\rho)^2$ for the four considered cases with fully periodic BCs. Each plotted value is normalized with the corresponding dissipative flux, {\it i.e.} the diffusivity and the viscous contributions for $\tilde{p}$ and $\phi$, respectively.}
\label{T2D_X}
\end{figure}

One can see how in all cases but the blue one the direction of the toroidal transitions for $\tilde{p}$ (left) depends on $k_\bot$, with a direct cascade at large wave-lengths ($T^{(n\neq0\rightarrow n=0)}_{\tilde{p}}<0$), partially compensating the source due to background pressure gradients (dashed lines), and an inverse cascade at large wave-lengths ($T^{(n\neq0\rightarrow n=0)}_{\tilde{p}}>0$). For $\phi$ (right), a direct cascade occurs in most of the spectral domain modulo some fluctuations in the two cases with classical diffusivity (yellow and green). A more complex pattern is shown by 2D transitions in Fig.\ref{T2D_X}, with essentially a direct poloidal cascade from very large wave-lengths for $\tilde{p}$, whose magnitude is of the same order as toroidal transitions, and strong oscillations around zero for $\phi$.  
  
\begin{figure}[h]
\centering
\includegraphics[width=\textwidth]{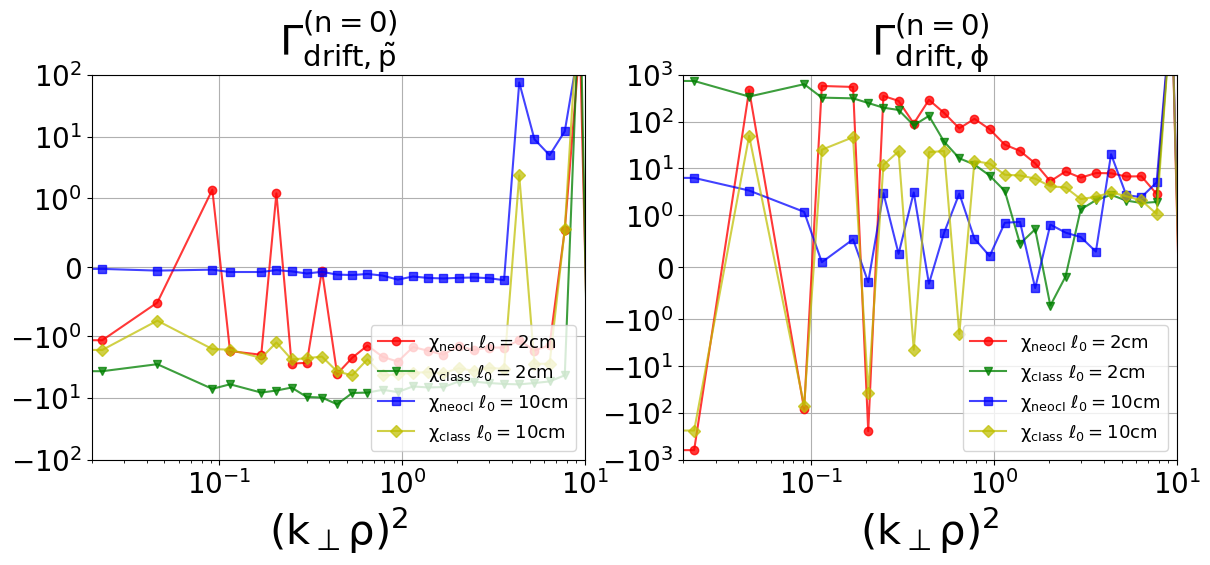}
\caption{The fluxes due to the drift coupling (solid lines) for $\tilde{p}$ (left) and $\phi$ (right) vs $(k_\bot\rho)^2$ for the four considered cases with fully periodic BCs. Each plotted value is normalized with the corresponding dissipative flux, {\it i.e.} the diffusivity and the viscous contributions for $\tilde{p}$ and $\phi$, respectively.}
\label{Drift_X}
\end{figure}

The drift contributions are shown in Fig.\ref{Drift_X}: they provide an energy flow from $\tilde{p}$ (left), for which they act as sink terms ($\Gamma_{\mathrm{drift, \tilde{p}}}<0$), to $\phi$ (right) where they mostly play the role of an energy source ($\Gamma_{\mathrm{drift, \phi}}>0$).  

The inverse toroidal cascade for $\tilde{p}$ at small poloidal wave-lengths means that non-axisymmetric modes combine to generate small scale axisymmetric modes. Hence, it can explain the observed correlation between non-axisymmetric modes and small vortex structures in the poloidal plane noted in the multi-panel figure Tab.\ref{frames_X} for $\tilde{p}$. Such a correlation extends to $\phi$ as well thanks to the drift term providing an energy flow from $\tilde{p}$ to $\phi$. 

The differences between the considered cases are outlined by the spectral analysis in time. This is done by computing the spectrogram, obtained by splitting the considered time window into overlapping intervals along which FFTs in time are computed and the mode square amplitude is shown. This procedure provides a visualization of the signal frequency content vs time. In what follows, we will fix a radial position below the X point ($x=1.64cm$) and plot the time average of the spectrogram at saturation while changing $y$ from SOL across PFR to SOL again but on the other side (we do not have here a distinction between LFS and HFS SOL since toroidicity is ignored, $B_\varphi=B_0=const.$). 

\begin{figure}[h]
\centering
\includegraphics[width=\textwidth]{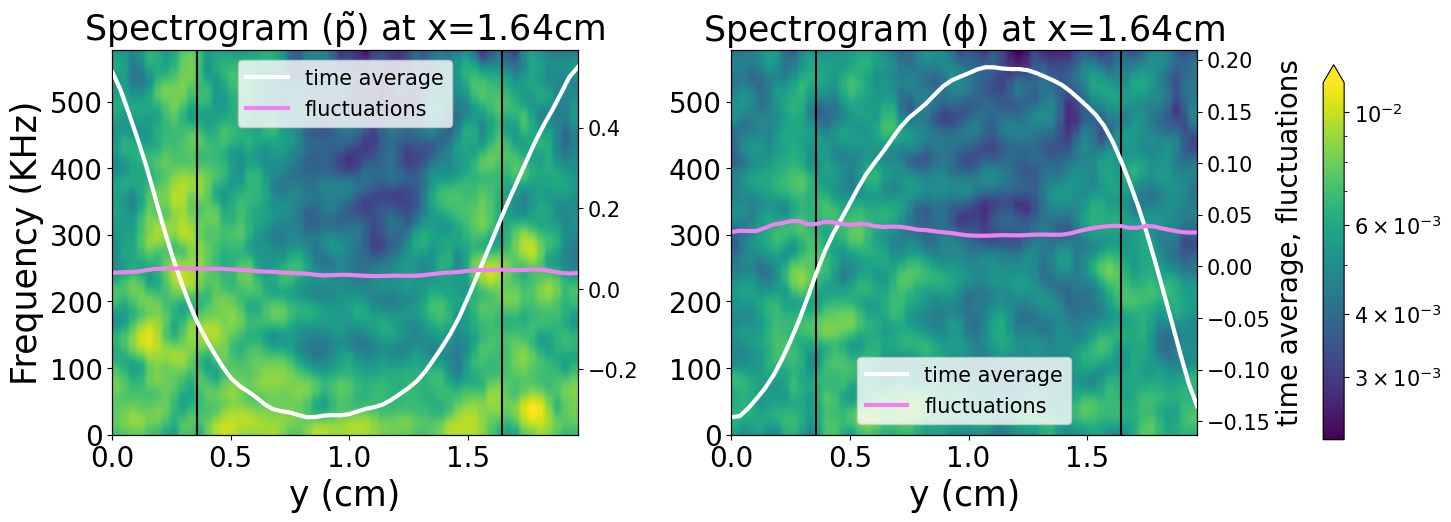}
\caption{The time average of the spectrograms for $\tilde{p}$ (left) and $\phi$ (right) in case b) $\chi_\bot=\chi_{neocl}$ and $\ell_0=2cm$ at $x=1.64cm$. They are computed along the same time window of the spectra in Fig.\ref{en_spectra_X} ($0.23 ms$) and single chunks are constructed through a Gaussian function of $7\mu s$ width with $1\mu s$ hop length.}
\label{spectr_l0x20_chix10}
\end{figure}

In case a) ($\chi_\bot=\chi_{neocl}$ and $\ell_0=2cm$) the spectrogram for $\tilde{p}$ (left) and $\phi$ (right) together with the corresponding time average (white curves) and fluctuations (violet curves) is shown in Fig.\ref{spectr_l0x20_chix10}. 
It is worth noting how the frequency spectrum is very wide, from zero up to the limit given by the time resolution of the output file ($\sim$500KHz). It is clearly visible a reduction of the spectral intensity in PFR, in correspondence to an under-dense region (the time average of $\tilde{p}<0$) and a peak of the electrostatic potential, while the fluctuations are nearly constant across all the considered poloidal region. 

\begin{figure}[h]
\centering
\includegraphics[width=\textwidth]{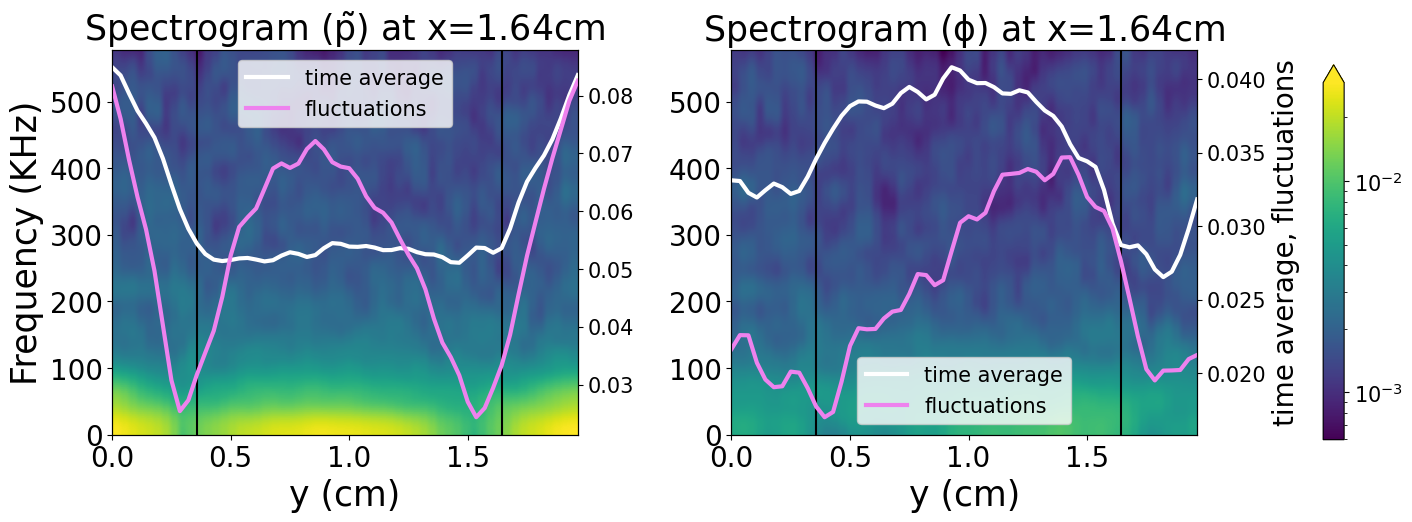}
\caption{The time average of the spectrograms for $\tilde{p}$ (left) and $\phi$ (right) in case a) $\chi_\bot=\chi_{class}$ and $\ell_0=2cm$ at $x=1.64cm$. They are computed along the same time window of the spectra in Fig.\ref{en_spectra_X} ($0.23 ms$) and single chunks are constructed through a Gaussian function of $7\mu s$ width with $1\mu s$ hop length.}
\label{spectr_l0x20_chi}
\end{figure}

The same quantities are shown in Fig.\ref{spectr_l0x20_chi} for the case b) with the same $\ell_0$ and classical $\chi_\bot$. 
The spectrogram has now a rapid decrease of the intensity above 100KHz. Moreover, one can notice two narrow quiescent regions for $\tilde{p}$ (left) to the left of the separatrix (black), one in SOL and the other in PFR, in correspondence to the minima of the fluctuations. A similar but less intense spectrogram is obtained in the other case (d) with classical diffusivity and it is not shown here, while the analysis in case c) is not so relevant since turbulence is quenched, as discussed before. 

Therefore, the spectral analysis in time outlines an unexpected feature of the turbulent spectra while changing diffusivity: the involved frequencies are much higher while increasing $\chi_\bot$. Moreover, for neoclassical diffusivity a quiescent region appears in PFR that seems to be due to the background dynamics, while in the classical case two narrow quiescent region are placed just to the left of the separatrix and they are characterized by a local suppression of the fluctuations. 

\subsection{Vanishing radial Dirichlet BCs}\label{sec3.2}

We present here the results of the simulations with vanishing radial Dirichlet BCs. Each panel of Tab.\ref{frames_X_NEW} corresponds to one of the four considered cases with strong/weak background pressure gradients for the top/bottom panels and neoclassical/classical diffusivity to the left/right. In each panel the fields profiles ($\tilde{p}$ to the left and $\phi$ to the right) are shown in the poloidal plane ($x,y$) (top figures) and in the radial-toroidal plane ($x,z$) (bottom figures). The full simulation domain is shown with the physical domain restricted to just the first half of the $x$ axis, while the second half contains specular copies of the field profiles. This means that only those modes corresponding to the sine transform are present and the vanishing Dirichlet boundary conditions has been correctly implemented.  

\begin{table}
\begin{tabular}{|c|c|}

      \hline \\
            \includegraphics[trim={3.5cm 1.5cm 3.5cm 1.7cm},clip,width=80mm]{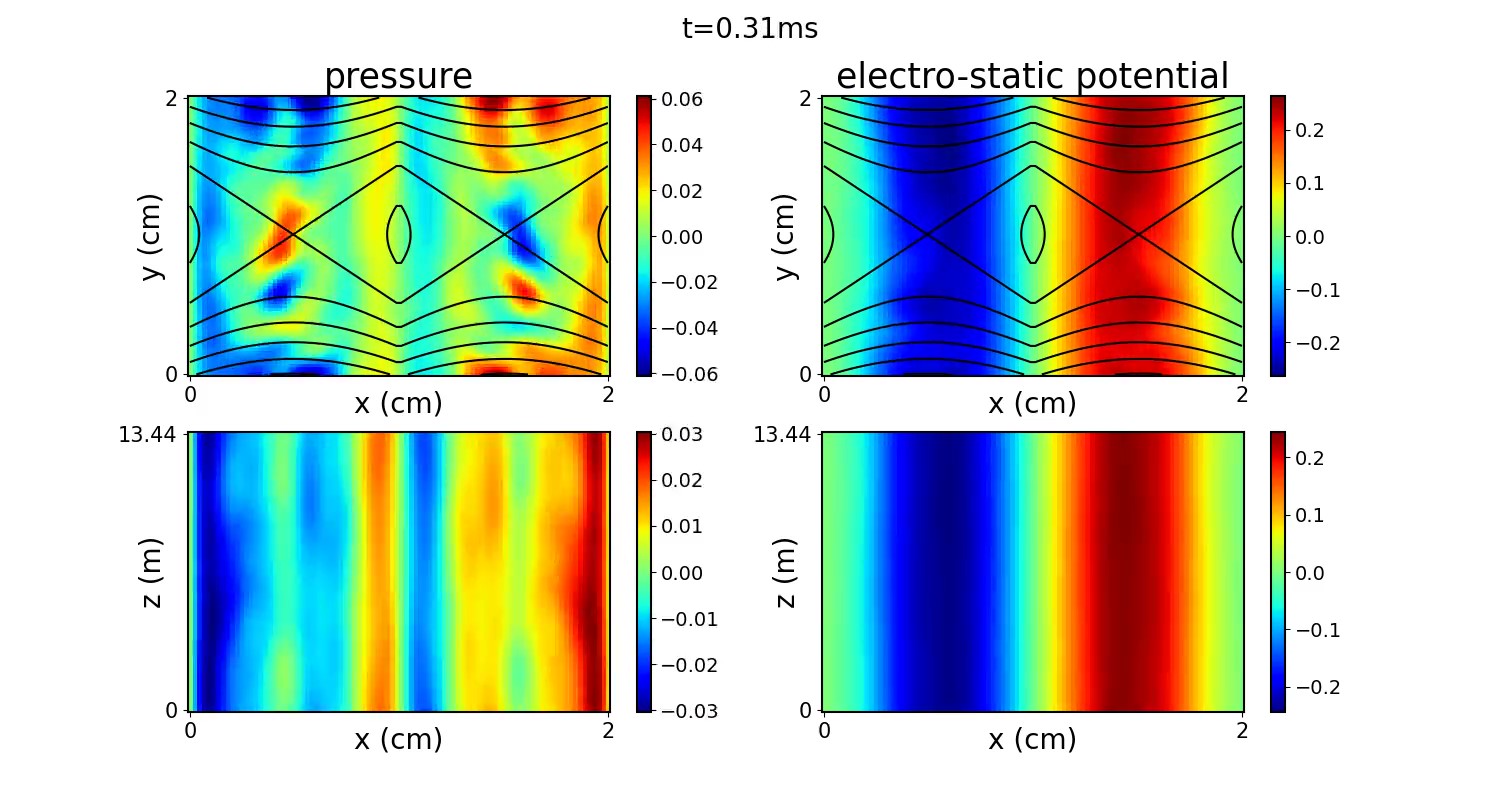} & \includegraphics[trim={3.5cm 1.5cm 3.5cm 1.7cm},clip,width=80mm]{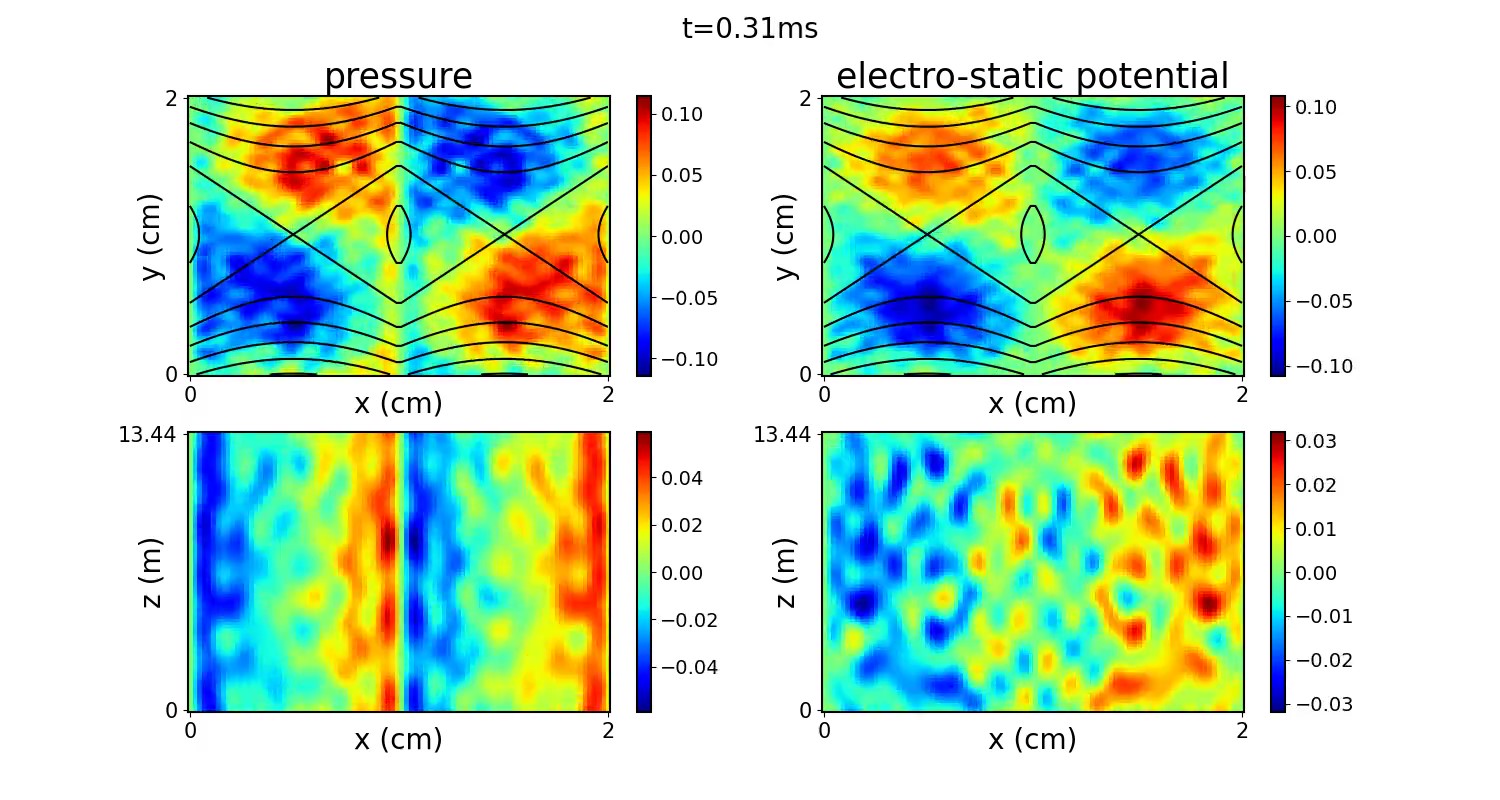} \\
			{\color{red}\textbullet} a) $\chi_{\mathrm{neocl}}$ $\ell_0$=2cm  & {\color{ForestGreen}\textbullet} b) $\chi_{\mathrm{class}}$ $\ell_0$=2cm \\
			\hline \\
      \includegraphics[trim={3.5cm 1.5cm 3.5cm 1.7cm},clip,width=80mm]{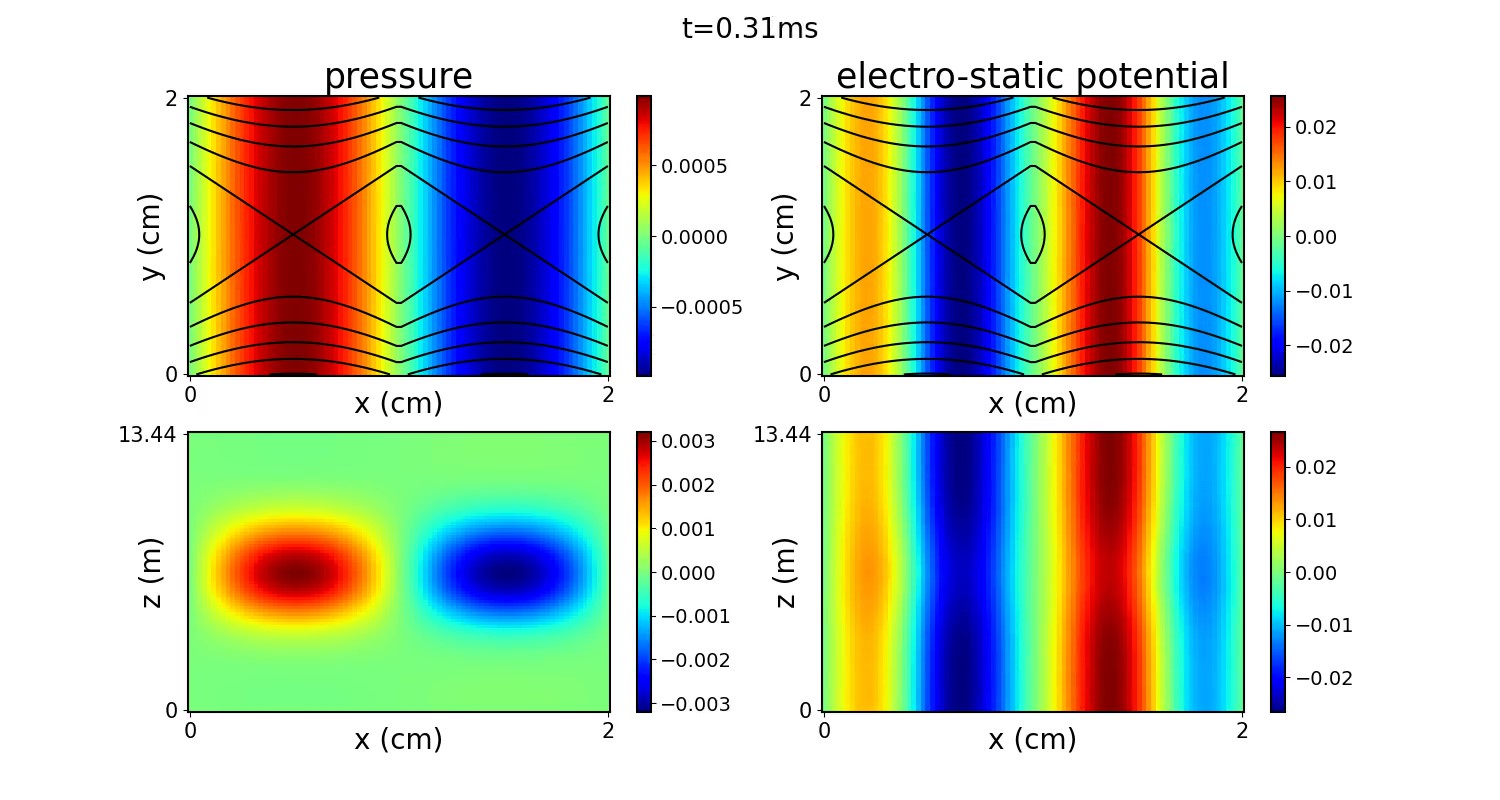} & \includegraphics[trim={3.5cm 1.5cm 3.5cm 1.7cm},clip,width=80mm]{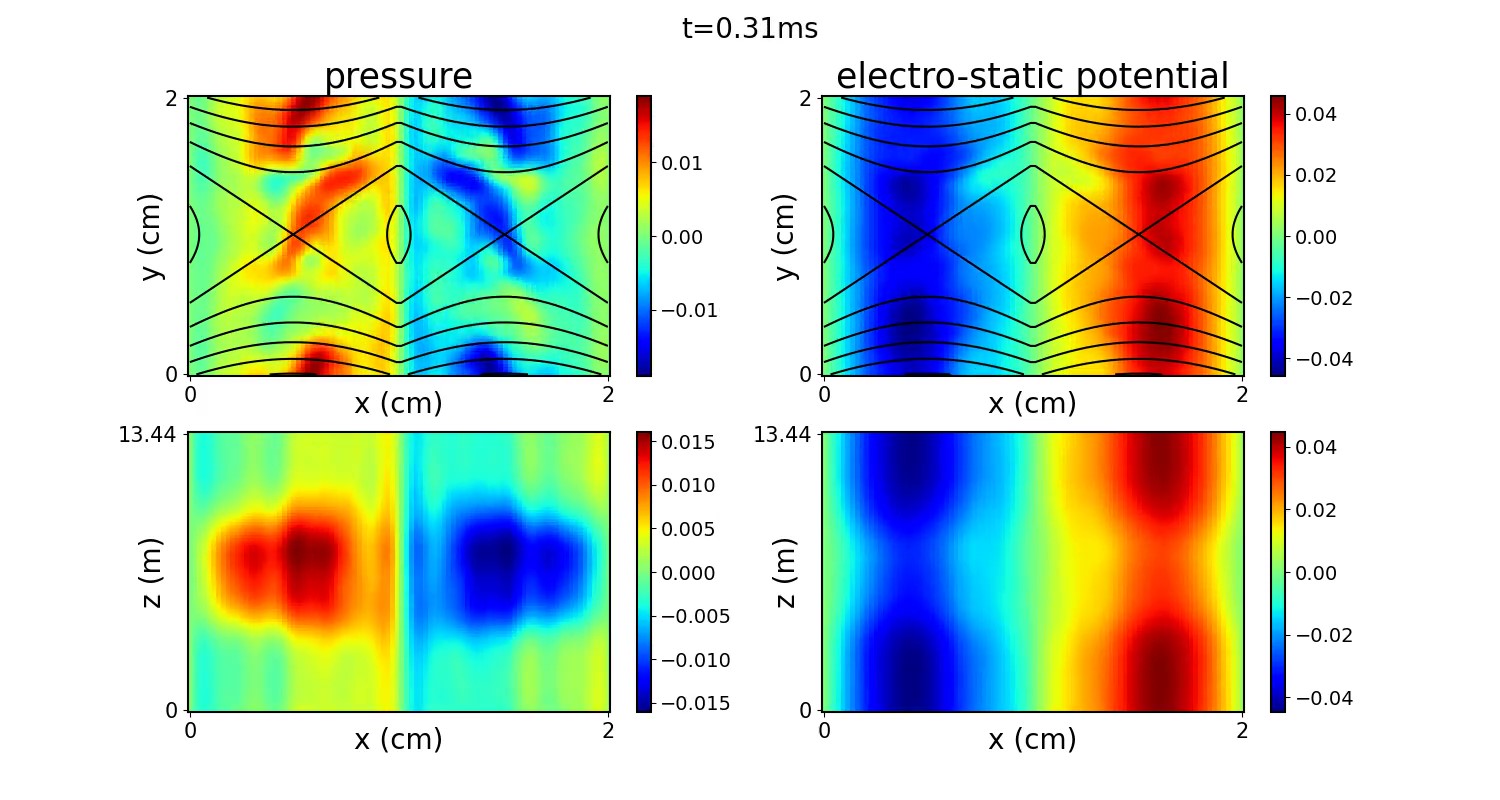} \\
			{\color{blue}\textbullet} c) $\chi_{\mathrm{neocl}}$ $\ell_0$=10cm & {\color{Goldenrod}\textbullet} d) $\chi_{\mathrm{class}}$ $\ell_0$=10cm \\			
			\hline 
\end{tabular}
\caption{The multi-panel figure showing the fields profile at fixed time $t=0.31ms$ for the simulations with vanishing radial Dirichlet BCs. Each panel is for one choice of the model parameters: a) $\chi_\bot=\chi_{\mathrm{neocl}}$ and $\ell_0$=2cm (top left), b) $\chi_\bot=\chi_{\mathrm{class}}$ and $\ell_0$=2cm  (top right), c) $\chi_\bot=\chi_{\mathrm{neocl}}$ and $\ell_0$=10cm (bottom left), d) $\chi_\bot=\chi_{\mathrm{class}}$ and $\ell_0$=10cm  (bottom right). In each panel the poloidal (top) and the radial-toroidal (bottom) sections of $\tilde{p}$ (left) and $\phi$ (right) are plotted with the corresponding colorbars next to each plot.}
\label{frames_X_NEW}
\end{table}

By comparing the panels in Tab.\ref{frames_X} with those in Tab.\ref{frames_X_NEW} it is clear how in the former case with fully periodic BCs the X magnetic geometry has a strong impact on the field profiles, while in the latter one sees that formation of elongated structures along $y$ and generically the obtained scenarios resemble those with a radially sheared poloidal field in Ref.\cite{Cianfrani:24}. This means that the imposition of the radial boundary conditions affects significantly the resulting drift turbulence scenario in X-point magnetic geometry. 
  
\begin{figure}[h]
\centering
\includegraphics[width=\textwidth]{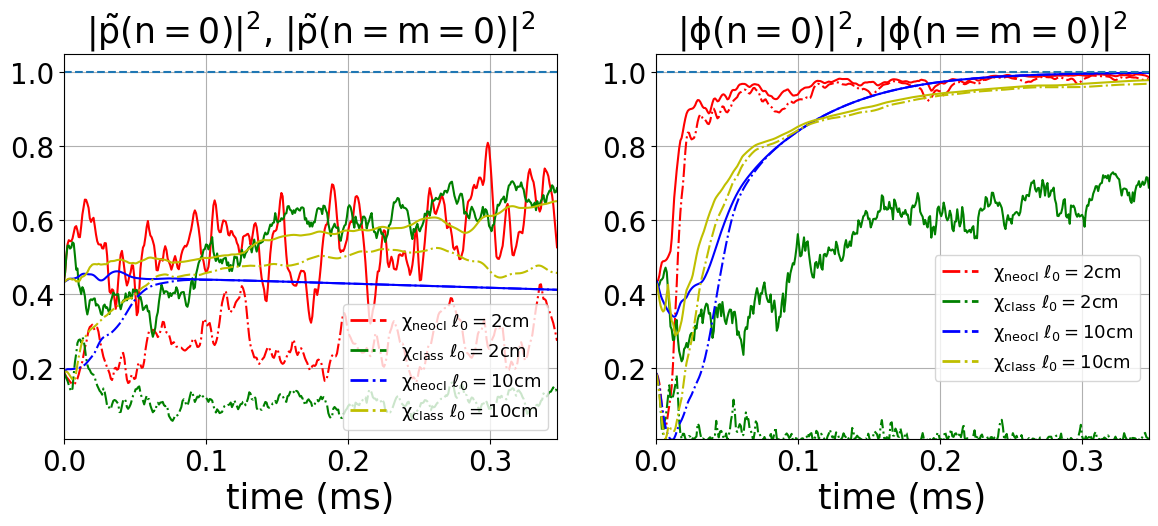}
\caption{The ratio of $n=0$ (solid lines) and of $n=m=0$ (dash-dotted lines) mode amplitudes over the total amplitude for $\tilde{p}$ (left) and $\phi$ (right) vs time (in ms) for the four considered cases with vanishing radial Dirichlet BCs.}
\label{modes_NEW_Xzf}
\end{figure}

A similar conclusion is obtained looking at the plots in Fig.\ref{modes_NEW_Xzf}, in which the ratio of $n=0$ (solid lines) and $n=m=0$ (dash-dotted lines) mode amplitudes to the total amplitude is shown in time, and comparing with those for radially sheared poloidal magnetic filed in Ref.\cite{Cianfrani:24} (Fig.25) that exhibit the same qualitative and to a large extent also quantitative behavior.

In particular, the electro-static potential (right) reaches saturated mode structures that are essentially toroidally and poloidally symmetric in all cases but b) (green), for which 60\% of mode amplitude is toroidally symmetric and less than 10\% of them are also poloidally symmetric. The scenario for $\tilde{p}$ (left) is very different: the fraction of $n=0$ modes is $\sim 60\%$ for cases a), b) and d) and more than 1/3 of them have also $m=0$. Case c) (blue) is peculiar as toroidally symmetric modes are also poloidally symmetric and their amplitude decays with time. 

\begin{figure}[h]
\centering
\includegraphics[width=\textwidth]{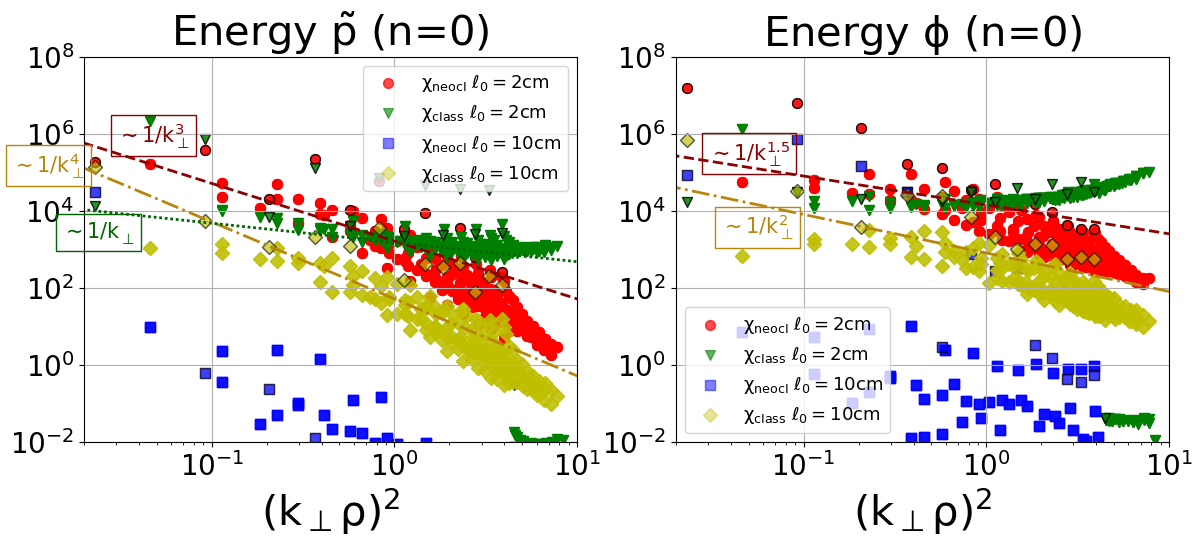}
\caption{The scatter plots of normalized axisymmetric spectral energies $E_{\tilde{p}}$ (left) and $E_\phi$ (right) vs $(k_\bot\rho)^2$ for the four considered cases with vanishing radial BCs. They are computed by averaging over a time window of $\sim 0.26ms$. Some curves are also plotted that approximate quite well some of the spectra profiles and which behave as: $1/k_\bot^3$ (dashed dark red line), $1/k_\bot^4$ (dash-dotted dark yellow line) and $1/k_\bot$ (dotted dark green line) for $E_{\tilde{p}}$ (left) and $1/k_\bot^{1.5}$ (dashed dark red line) and $1/k_\bot^2$ (dash-dotted dark yellow line) for $E_\phi$ (right).}
\label{en_spectra_NEW_X}
\end{figure}

A clear distinction with the fully periodic BCs discussed above (see Fig.\ref{en_spectra_X}) is found in the 2D anisotropic spectra in Fig.\ref{en_spectra_NEW_X} characterized by the enhancement of poloidally symmetric $m=0$ modes (black edges). 
It is also quite impressive the similarity with the spectra for a radially sheared poloidal magnetic field in Ref.\cite{Cianfrani:24} (see Fig.26). 

\begin{figure}[h]
\centering
\includegraphics[width=\textwidth]{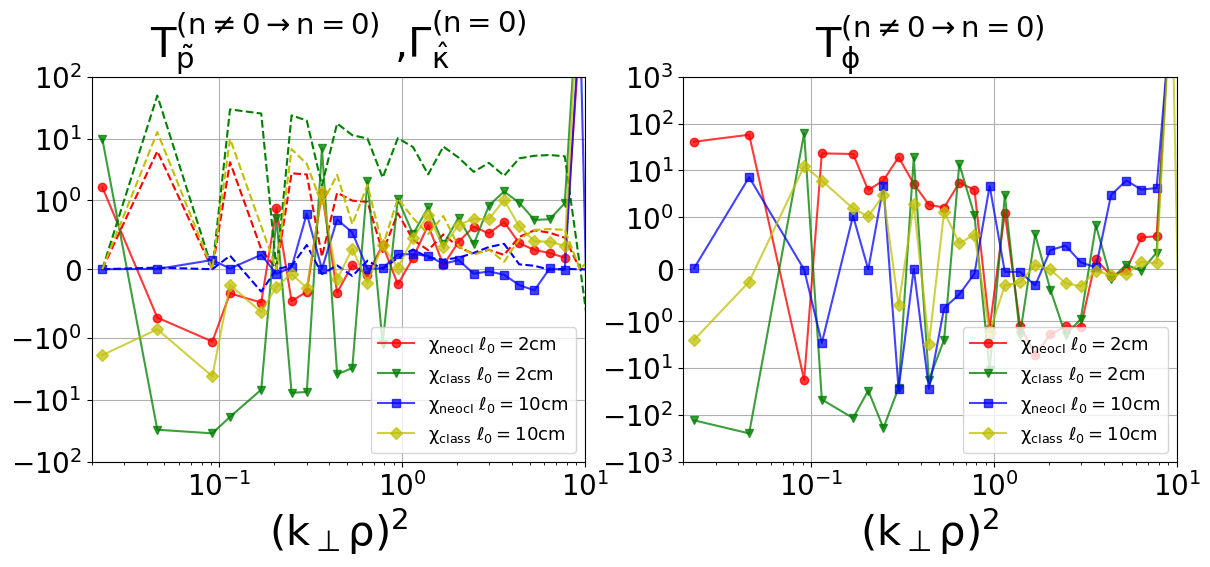}
\caption{The nonlinear transfer functions of toroidal transitions $n\neq0\rightarrow n=0$ (solid lines) for $\tilde{p}$ (left) and $\phi$ (right) vs $(k_\bot\rho)^2$ and the flux due to background pressure gradient (dashed lines, left) for the four considered cases with vanishing radial Dirichlet BCs. Each plotted value is normalized with the corresponding dissipative flux, {\it i.e.} the diffusivity and the viscous contributions for $\tilde{p}$ and $\phi$, respectively. The chosen values of $(k_\bot\rho)^2$ are constructed from the computational Cartesian grid in FS by taking the five smallest values of $k_\bot$ and by logarithmically sampling the relic part.}
\label{Tpphi_NEW_X}
\end{figure}

\begin{figure}[h]
\centering
\includegraphics[width=\textwidth]{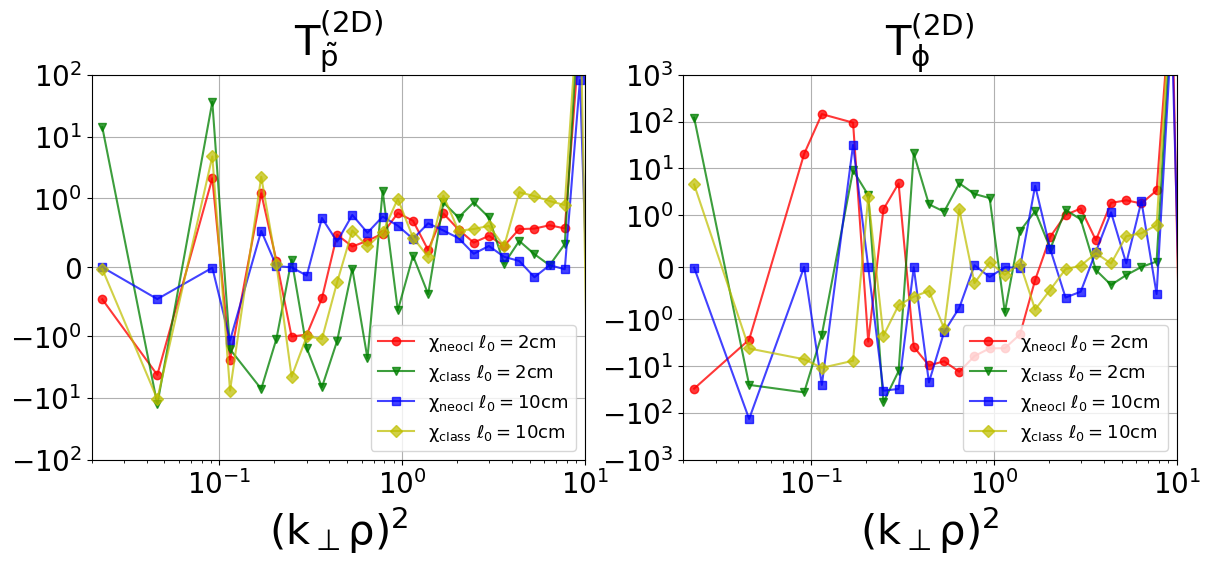}
\caption{The nonlinear transfer functions of 2D transitions $n=0\rightarrow n=0$ (solid lines) for $\tilde{p}$ (left) and $\phi$ (right) vs $(k_\bot\rho)^2$ for the four considered cases with vanishing radial Dirichlet BCs. Each plotted value is normalized with the corresponding dissipative flux, {\it i.e.} the diffusivity and the viscous contributions for $\tilde{p}$ and $\phi$, respectively.}
\label{T2D_NEW_X}
\end{figure}

\begin{figure}[h]
\centering
\includegraphics[width=\textwidth]{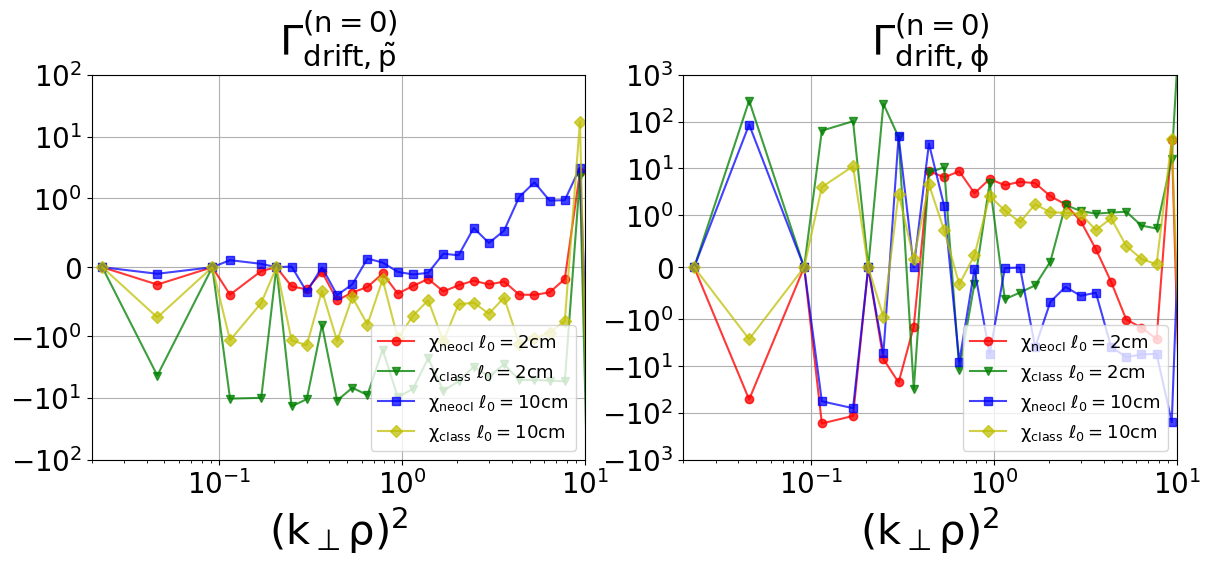}
\caption{The fluxes due to the drift coupling (solid lines) for $\tilde{p}$ (left) and $\phi$ (right) vs $(k_\bot\rho)^2$ for the four considered cases with vanishing radial Dirichlet BCs. Each plotted value is normalized with the corresponding dissipative flux, {\it i.e.} the diffusivity and the viscous contributions for $\tilde{p}$ and $\phi$, respectively.}
\label{Drift_NEW_X}
\end{figure}

Fluxes have the same qualitative spectral features as in the case with fully periodic boundary conditions:
\begin{itemize}
\item the toroidal transitons for $\tilde{p}$ provide a direct cascade at large wave-lengths and an inverse cascade at small wave-lengths (see Fig.\ref{Tpphi_NEW_X}),
\item the fluxes due to 2D transitions (see Fig.\ref{T2D_NEW_X}) are of the same order as those for toroidal ones,
\item the drift terms (see Fig.\ref{Drift_NEW_X}) tend to drive energy from $\tilde{p}$ to $\phi$.  
\end{itemize}

The spectral analysis in time does not give any significant insight into the turbulent scenarios, with the spectrograms being suppressed above 100kHZ similarly to that in Fig.\ref{spectr_l0x20_chi} but without any quiescent region. 

Therefore, the impact of the X-point magnetic configuration is greatly reduced for the simulations with vanishing radial Dirichlet BCs and the resulting spectral features resemble those of the radially sheared magnetic field discussed in Ref.\cite{Cianfrani:24}. The reason for that could be ascribed to the leading role of $m=0$ modes, such that the field profiles are insensitive to those background structure that are not constant in $y$, such as the radial component of the magnetic field $B_x\propto y$ (see eq.(\ref{mag_field})), and they effectively see just the radially sheared poloidal magnetic field component.    

\section{Conclusions}\label{sec4}
We presented electro-static simulations of edge tokamak plasma for a DTT-like scenario near the X point in the low-frequency limit, taking the electro-static potential $\phi$ and the perturbed pressure $\tilde{p}$ as field variables. In order to interpret the results, we made a comparison with the outcomes of the companion paper Ref.\cite{Cianfrani:24} where a similar analysis is performed with simpler magnetic configurations, namely SLAB, a constant and a radially sheared poloidal field in addition to the toroidal component.   

We discussed the relevance of two model parameters for the emerging turbulent scenario. These are the background pressure-gradient length scale $\ell_0$ fixing the free-energy source of the system and perpendicular diffusivity $\chi_\bot$, that determines the amount of dissipation which is complementary to the drift response. We found the same qualitative results as in Ref.\cite{Cianfrani:24}, {\it i.e.} a correlation between increasing $1/\ell_0$ and/or decreasing $\chi_\bot$ with the emergence of toroidal asymmetries and small scale structures in the poloidal plane. The analysis of fluxes outlined the crucial role of toroidal transitions, thus the necessity of 3D simulations, for the resulting saturated spectra even though the leading modes are axisymmetric ones ($n=0$). In fact, looking at $\tilde{p}$ energy flow a direct toroidal cascade at large wave-lengths co-exists with an inverse toroidal one at small wave-lengths. This finding explains the connection between toroidal asymmetries and small scale poloidal structures for $\tilde{p}$. The same connection extends also to $\phi$ due to the drift term, that acts as a sink for $\tilde{p}$ and as a source for $\phi$, thus transferring energy between the two fields preserving the wave-length. 

The combination of all these effects with other contributions (such as the mostly direct toroidal cascade for $\phi$ and 2D transitions) provide the saturated 2D spectral profiles (see Fig.\ref{en_spectra_X}) that are quite isotropic in the case with fully periodic boundary conditions. This achievement must be compared with the strong anisotropy obtained in those cases with a constant or sheared poloidal magnetic field component discussed in Ref.\cite{Cianfrani:24}, in which the enhancement of poloidally symmetric $m=0$ modes (zonal flow) was predicted. The different outcomes can be ascribed to the formal symmetry of the drift term near the X point under exchange of the coordinates $x$ and $y$ in the poloidal plane. Therefore, the X point morphology does not induce the excitation of zonal flows, while the more conventional magnetic geometry with just a poloidal field component in addition to the toroidal one is expected to generate them, posing the fundamental question of how the two mechanisms can co-exist in a global framework and how they affect the propagation of macroscopic structures (filaments, blobs).   

A spectral analysis in time (spectrograms) has also been performed along a line below the X-point and it showed the different characteristics of the turbulent modes when changing $\chi_\bot$: in the case with neoclassical diffusivity a wide frequency range is covered, up to 500KHzs, and we noticed a quiescent PFR, while for classical diffusivity those modes with frequencies above 100KHz are suppressed and two narrow quiescent regions are predicted near the separatrix legs, one in PFR and the other in SOL. The nature of these quiescent regions appear also very different, with the latter due to a local reduction of the fluctuations level, while the former is correlated with the formation of an underdense region and of a local maximum of the potential.   

The imposition of vanishing Dirichlet BCs along the radial direction $x$ provides significant differences compared to those simulations with fully periodic boundary conditions. In particular, the spectra are not anymore anisotropic with poloidally symmetric $m=0$ modes being enhanced than the other modes with the same $k_\bot$ (see the energy spectra in Fig.\ref{en_spectra_X}). Moreover, the results are very close to those obtained with a radially sheared poloidal magnetic field in Ref.\cite{Cianfrani:24}. This is an indication of the prominent role of the radial BCs in suppressing $m\neq 0$ modes to the extent that the background $y$-dependent magnetic structure, {\it i.e.} the radial component $B_x$, is not relevant at all and the fields ``see'' an effective magnetic geometry with just a radially sheared poloidal field $B_y$. Therefore, the choice of the BCs is crucial for the spectral features of the emerging turbulent scenario near the X point magnetic geometry and the investigation of more general conditions (sheat BCs) will be the subject of further developments of this approach.    

This paper and the companion one \cite{Cianfrani:24} provide a good understanding of the main physical picture behind drift turbulence in the edge of a tokamak plasma near the X point in terms of dependence on the model parameters, energy fluxes, relevance of the toroidal transitions, excitation of zonal flows, role of BCs. The inclusion of some additional terms accounting for the curvature contribution, electro-magnetic fluctuations, diamagnetic velocity or gyrofluid corrections will be performed straightforwardly by updating the simulation code. Similarly, future investigations will focus to other magnetic configurations (snowflake, Super-X, Double Super-X) that can be realized in forthcoming Italian DTT machine \cite{DTT3}. 

However, the main drawback of this approach is its locality in space, while global effects are expected to be relevant for the description of turbulent transport in a reactor. Therefore, in parallel some technical and modelization efforts will be devoted to the extension of the presented analysis to poloidally global simulations that could give us a realistic picture on the development of macroscopic structures (zonal flows, blobs and filaments) in view of the comparison with the achievement of those advanced codes \cite{Shanahan:14,Halpern:16,Tamain:16,Stegmeier:18} focused on reproducing the experimental conditions.     
 

\end{document}